\documentclass[12pt]{article}
\usepackage[dvips]{graphics,color}
\usepackage{latexsym,amssymb}
\usepackage{amsmath,amsbsy}
\usepackage{amsopn,amstext}
\usepackage{cite}
\usepackage{amssymb}
\usepackage{rotating}
\usepackage{amsmath}

\allowdisplaybreaks[4]

\renewcommand{\theequation}{\arabic{section}.\arabic{equation}}

\begin{document}

%************************** Text Begins here ******************************

%  Greek letters

\def\a{\alpha}
\def\b{\beta}
\def\d{\delta}
\def\e{\epsilon}
\def\g{\gamma}
\def\h{\mathfrak{h}}
\def\k{\kappa}
\def\l{\lambda}
\def\o{\omega}
\def\p{\wp}
\def\r{\rho}
\def\t{\tau}
\def\s{\sigma}
\def\z{\zeta}
\def\x{\xi}
\def\V={{{\bf\rm{V}}}}
 \def\A{{\cal{A}}}
 \def\B{{\cal{B}}}
 \def\C{{\cal{C}}}
 \def\D{{\cal{D}}}
\def\G{\Gamma}
\def\K{{\cal{K}}}
\def\O{\Omega}
\def\R{\bar{R}}
\def\T{{\cal{T}}}
\def\L{\Lambda}
\def\f{E_{\tau,\eta}(sl_2)}
\def\E{E_{\tau,\eta}(sl_n)}
\def\Zb{\mathbb{Z}}
\def\Cb{\mathbb{C}}

\def\R{\overline{R}}
% Shorthands for \begin{equation} and the like

\def\beq{\begin{equation}}
\def\eeq{\end{equation}}
\def\bea{\begin{eqnarray}}
\def\eea{\end{eqnarray}}
\def\ba{\begin{array}}
\def\ea{\end{array}}
\def\no{\nonumber}
\def\le{\langle}
\def\re{\rangle}
\def\lt{\left}
\def\rt{\right}

\newtheorem{Theorem}{Theorem}
\newtheorem{Definition}{Definition}
\newtheorem{Proposition}{Proposition}
\newtheorem{Lemma}{Lemma}
\newtheorem{Corollary}{Corollary}
\newcommand{\proof}[1]{{\bf Proof. }
        #1\begin{flushright}$\Box$\end{flushright}}

\baselineskip=20pt

%%%%%%%%%%%%%%%%%%%%%%%%%%%%%%%%%%%%%%%%%%%%%%%%%%%%%%%%%%%%
%                                                          %
%  Title page                                              %
%                                                          %
%%%%%%%%%%%%%%%%%%%%%%%%%%%%%%%%%%%%%%%%%%%%%%%%%%%%%%%%%%%%
\newfont{\elevenmib}{cmmib10 scaled\magstep1}
\newcommand{\preprint}{
   \begin{flushleft}
     %\elevenmib Yukawa\, Institute\, Kyoto\\
   \end{flushleft}\vspace{-1.3cm}
   \begin{flushright}\normalsize
  % \sf  YITP-03-53\\
   %  {\tt hep-th/yymmnnn} \\ November 2005
   \end{flushright}}
\newcommand{\Title}[1]{{\baselineskip=26pt
   \begin{center} \Large \bf #1 \\ \ \\ \end{center}}}

\newcommand{\Author}{\begin{center}
   \large \bf
Guang-Liang Li${}^{a}$, Junpeng Cao${}^{b,c,d}$, Panpan
Xue${}^{a}$,  Kun Hao${}^{e,f}$, Pei Sun${}^{e,f}$, Wen-Li
Yang${}^{e,f,g}\footnote{Corresponding author: wlyang@nwu.edu.cn}$,
Kangjie Shi${}^{e,f}$ and Yupeng
Wang${}^{b,d,h}\footnote{Corresponding author: yupeng@iphy.ac.cn}$
 \end{center}}

\newcommand{\Address}{\begin{center}

     ${}^a$Department of Applied Physics, Xian Jiaotong University, Xian 710049, China\\
     ${}^b$Beijing National Laboratory for Condensed Matter
           Physics, Institute of Physics, Chinese Academy of Sciences, Beijing
           100190, China\\
     ${}^c$School of Physical Sciences, University of Chinese Academy of
Sciences, Beijing, China\\
     ${}^d$Songshan Lake Materials Laboratory, Dongguan, Guangdong 523808, China \\
     ${}^e$Institute of Modern Physics, Northwest University,
     Xian 710127, China\\
     ${}^f$ Shaanxi Key Laboratory for Theoretical Physics Frontiers,  Xian 710127, China\\
     ${}^g$ School of Physics, Northwest University,  Xian 710127, China\\
     ${}^h$ The Yangtze River Delta Physics Research Center, Liyang, Jiangsu, China

\end{center}}

\newcommand{\Accepted}[1]{\begin{center}
   {\large \sf #1}\\ \vspace{1mm}{\small \sf Accepted for Publication}
   \end{center}}

\preprint
\thispagestyle{empty}
\bigskip\bigskip\bigskip

\Title{Off-diagonal Bethe Ansatz on the $so(5)$ spin chain} \Author

\Address
\vspace{1cm}

\begin{abstract}
The $so(5)$ (i.e., $B_2$) quantum integrable spin chains with both periodic and non-diagonal boundaries are studied via the off-diagonal Bethe Ansatz method. By using the fusion technique, sufficient operator product identities (comparing to those in \cite{NYReshetikhin1}) to determine the spectrum of the transfer matrices are derived. For the periodic case, we recover the results obtained in \cite{NYReshetikhin1}, while for the non-diagonal boundary case, a new inhomogeneous $T-Q$ relation is constructed. The present method can be directly generalized to deal with the $so(2n+1)$ (i.e., $B_n$) quantum integrable spin chains with general boundaries.

\vspace{1truecm} \noindent {\it PACS:} 75.10.Pq, 02.30.Ik, 71.10.Pm

\end{abstract}
\newpage
%%%%%%%%%%%%%%%%%%%%%%%%%%%%%%%%%%%%%%%%%%%%%%%%%%%%%%%%%%%%%%%
%                                                             %
%  1. Introduction                                            %
%                                                             %
%%%%%%%%%%%%%%%%%%%%%%%%%%%%%%%%%%%%%%%%%%%%%%%%%%%%%%%%%%%%%%%

\section{Introduction}

Both the algebraic and coordinate Bethe Ansatz are very powerful methods to construct exact solutions of quantum
integrable models\cite{1,2,Tak79,Kor93,Alc87,Skl88}. Nevertheless, those methods still have their restrictions for depending on existence of an obvious reference state. An important issue is that when $U(1)$-symmetry is broken, the systems may not have obvious reference states. In such cases, the problem becomes more frustrated and  many interesting efforts have been made in this direction \cite{Fan96,Nep04,Cao03,Gie04,Yan04,Gie05,Bas1,Bas2,Gal08,Fra08,Bas3,Nic12,Cao13-1,Nep13,Bel13,Bel15,Bel15-1,Ava15} in the past several decades. A generic method named off-diagonal Bethe ansatz (ODBA) for solving quantum integrable models either with or without $U(1)$-symmetry was proposed in \cite{Cao1}. By constructing the inhomogeneous $T-Q$ relations based on some operator identities, several typical models are solved exactly \cite{Cao2}. With the resulting
eigenvalues, the corresponding Bethe-type eigenstates can also be retrieved\cite{Zha13,Hao14}. The nested ODBA was initially proposed in studying the $su(n)$ (i.e., $A_n$) spin chain with generic boundaries \cite{Cao14JHEP143,Cao15JHEP036}. However, ODBA to approach high-rank quantum integrable models associated
with $B_n$, $C_n$ and $D_n$ Lie algebras is still missing. We note that such kind of models with obvious $U(1)$-symmetry has been studied extensively. For example, with some functional relations and algebraic Bethe Ansatz analysis (the analytic Bethe Ansatz method), Reshetikhin derived the energy spectrum of the periodic quantum spin chains associated with $B_n$, $C_n$, $D_n$ and other Lie algebras \cite{NYReshetikhin1,NYReshetikhin2}.
The algebraic Bethe Ansatz for those models with periodic boundary condition was constructed by Martins and Ramos \cite{Bn}, while the method for approaching such kind of models with diagonal open boundaries was developed by
Li, Shi and Yue \cite{b2aba,c2aba}.

In this paper, we develop a nested ODBA method to approach the quantum integrable $so(5)$ (i.e., $B_2$) spin chain with either periodic or non-diagonal open boundary condition. This method  can be generalized to $so(2n+1)$ (i.e., $B_n$) case directly. The paper is organized as follows. In section 2, we study the $so(5)$ model with periodic boundary condition.
Closed functional relations among the transfer matrices to determine the eigenvalues are constructed with fusion techniques.
In section 3, we study the $so(5)$ model with an off-diagonal open boundary condition.
By constructing some operator product identities, we derive the exact eigenvalues of the transfer matrix in terms of an inhomogeneous $T-Q$ relation. Section 4 is attributed to concluding remarks. Some detailed calculations are listed in Appendices A-C.

\section{$so(5)$ spin chain with periodic boundary condition}
\setcounter{equation}{0}

\subsection{The model}

Let  ${\rm V}$ denote a $5$-dimensional linear space with an orthonormal basis $\{|i\rangle|i=1,\cdots,5\}$  which endows the fundamental
representation of the $so(5)$ (or $B_2$) algebra.
The quantum spin chain associated with the $B_2$ algebra is described by a $25\times 25$ $R$-matrix $R^{vv}_{12}(u)$ defined in the $V \otimes V$ space with the matrix elements \cite{Bn}
\begin{eqnarray}
   R_{12}^{  vv}(u)^{ij}_{kl} = u(u+\frac{3}{2})\delta_{ik}\delta_{jl}+
(u+\frac{3}{2})\delta_{il}\delta_{jk}-u\delta_{j\bar{i}}\delta_{k\bar{l}},
 \label{rm}
\end{eqnarray}
where  $\{i,j,k,l\}=\{1,2,3,4,5\}$, $i+\bar{i}=6$.
We introduce the notation for simplicity
\begin{eqnarray}
&&R_{12}^{  vv}(u)^{ii}_{ii}=a_1(u)=(1+u)(u+\frac{3}{2}),\quad i\ne 3, \nonumber\\[4pt]
&&R_{12}^{  vv}(u)^{ij}_{ij}=b_1(u)=u(u+\frac{3}{2}),\quad i\ne j, \bar{j}, \nonumber\\[4pt]
&&R_{12}^{  vv}(u)^{i\bar{i}}_{\bar{i}i}=c_1(u)=\frac{3}{2},\quad i\ne \bar{i},\nonumber\\[4pt]
&&R_{12}^{  vv}(u)^{i\bar{i}}_{j\bar{j}}=d_1(u)=-u,\quad i\ne j, \bar{j},\nonumber\\[4pt]
&&R_{12}^{  vv}(u)^{i\bar{i}}_{i\bar{i}}=e_1(u)=u(u+\frac{1}{2}),\quad i\ne \bar{i}, \nonumber\\[4pt]
&&R_{12}^{  vv}(u)^{ii}_{ii}=f_1(u)=a_1(u)+d_1(u),\quad i= 3, \nonumber\\[4pt]
&&R_{12}^{  vv}(u)^{ij}_{j{i}}=g_1(u)=u+\frac{3}{2}, \quad i\ne j, \bar{j}.
 \label{2}
\end{eqnarray}
The $R$-matrix satisfies the properties \cite{Nep-18}
\begin{eqnarray}
\hspace{-0.8truecm}{\rm regularity}&:&R^{  vv}_{12}(0)=\rho_1(0)^{\frac{1}{2}}{\cal P}_{12},\nonumber\\[4pt]
\hspace{-0.8truecm}{\rm unitarity}&:&R^{   vv}_{12}(u)R^{  vv}_{21}(-u)=\rho_1(u)=a_1(u)a_1(-u),\nonumber\\[4pt]
\hspace{-0.8truecm}{\rm crossing-symmetry}&:&R^{  vv}_{12}(u)=V_1\,\{R^{
vv}_{12}(-u-\frac{3}{2})\}^{t_2}\,V_1=V_2\,\{R^{
vv}_{12}(-u-\frac{3}{2})\}^{t_1}\,V_2,\label{Crossing-symmetry}
\end{eqnarray}
where ${\cal P}_{12}$ is the permutation operator with the matrix elements $[{\cal
P}_{12}]^{ij}_{kl}=\delta_{il}\delta_{jk}$, $t_i$ denotes the
transposition in the $i$-th space, $R _{21}={\cal
P}_{12}R _{12}{\cal P}_{12}$, and the crossing-matrix $V$ is
\begin{eqnarray}
V=\lt(\begin{array}{ccccc}&&&&1\\
&&&1&\\&&1&&\\&1&&&\\1&&&&
\end{array}\rt),\quad V^2={\rm id}.
\end{eqnarray}
Combining  the crossing-symmetry and the unitarity of the $R$-matrix, one can derive the relation
\begin{eqnarray}
{\rm crossing-unitarity}&:&R^{  vv}_{12}(u)^{t_1}R^{
vv}_{21}(-u-3)^{t_1}=\rho_1(u+\frac{3}{2}).\label{Permutation-1}
\end{eqnarray}
Here and below we adopt the standard notation: for any
matrix $A\in {\rm End}({\rm V})$, $A_j$ is an embedding operator
in the tensor space ${\rm V}\otimes {\rm V}\otimes\cdots$,
which acts as $A$ on the $j$-th space and as an identity on the
other factor spaces; $R_{ij}(u)$ is an embedding operator of
$R$-matrix in the tensor space, which acts as an identity on the
factor spaces except for the $i$-th and $j$-th ones.
The $R$-matrix satisfies the Yang-Baxter equation
\begin{eqnarray}
R^{   vv}_{12}(u-v)R^{  vv}_{13}(u)R^{    vv}_{23}(v)=R^{
  vv}_{23}(v)R^{   vv}_{13}(u)R^{  vv}_{12}(u-v). \no
\end{eqnarray}

For the periodic boundary condition, we introduce the monodromy matrix
\bea
T_0^{  v}(u)=R^{    vv}_{01}(u-\theta_1)R^{
vv}_{02}(u-\theta_2)\cdots R^{
   vv}_{0N}(u-\theta_N), \label{Mon-1}
\eea
where the index $0$ indicates the auxiliary space and the other tensor space  $V^{\otimes N}$ is the physical or quantum space, $N$ is the number of sites and $\{\theta_j\}$ are the inhomogeneous parameters.
The monodromy matrix satisfies the Yang-Baxter relation
\bea
 R^{  vv}_{12}(u-v) T_1^{  v}(u) T_2^{  v}(v) = T_2^{  v}(v) T_1^{  v}(u) R^{  vv}_{12}(u-v).\label{ybta2o}
\eea
The transfer matrix is the trace of monodromy matrix in the auxiliary space
\bea t^{(p)}(u)\stackrel{\rm def}{=}t_1^{(p)}(u)=tr_0 T_0^{  v}(u). \label{1117-1}\eea
From the Yang-Baxter relation, one can prove that the transfer matrices with different spectral parameters
commute with each other, $[t^{(p)}(u), t^{(p)}(v)]=0$. Therefore, $t^{(p)}(u)$ serves
as the generating function of all the conserved quantities of the
system. The Hamiltonian is given  by
\begin{eqnarray}
H_p= \frac{\partial \ln t^{(p)}(u)}{\partial
u}|_{u=0,\{\theta_j\}=0}.
\end{eqnarray}

\subsection{Spinorial $R$-matrix and the fused ones}
In order to obtain closed operator product identities (see (\ref{Op-Product-Periodic-1})-(\ref{futp-7}) below) which allow one to completely determine the eigenvalues of the  transfer matrix $t^{(p)}(u)$, we need further an $R$-matrix associated with the spinorial representation of  the $so(5)$ algebra. Let us  denote the spinorial representation   by ${\rm V^{(s)}}$ with  an orthonormal basis $\{|i\rangle^{(s)}|i=1,\cdots,4\}$. The spinorial $16\times 16$ $R$-matrix has the following non-zero matrix elements \cite{spm}
\begin{eqnarray}
&&R_{12}^{ ss}(u)^{ii}_{ii}=a_2(u)=(u+\frac{1}{2})(u+\frac{3}{2}),\nonumber\\
&&R_{12}^{  ss}(u)^{ij}_{ij}=b_2(u)=u(u+\frac{3}{2}),\quad i\ne j, \bar{j},\nonumber\\
&&R_{12}^{  ss}(u)^{i\bar{i}}_{\bar{i}i}=c_2(u)=u+\frac{3}{4},\nonumber\\
&&\xi_i\xi_{\bar j}R_{12}^{  ss}(u)^{i\bar{i}}_{j\bar{j}}=d_2(u)=-\frac{u}{2},\quad i\ne j, \bar{j},\nonumber\\
&&R_{12}^{  ss}(u)^{i\bar{i}}_{i\bar{i}}=e_2(u)=u(u+1),\nonumber\\
&&R_{12}^{  ss}(u)^{ij}_{j{i}}=g_2(u)=\frac{u}{2}+\frac{3}{4}, \quad i\ne j, \bar{j},
\label{Rss-matrix}
\end{eqnarray}
where $\{i,j\}=\{1,2,3,4\}$, $i+\bar{i}=5$,
$\xi_i=1$ if $ i\in \{1,2\}$ and $\xi_i=-1$ if $ i\in \{3,4\}$.
The spinorial $R$-matrix satisfies the properties
\begin{eqnarray}
{\rm regularity}&:&R^{  ss}_{12}(0)=\rho_2(0)^{\frac{1}{2}}{\cal P}^{(s)}_{12},\nonumber\\
{\rm unitarity}&:&R^{  ss}_{12}(u)R^{  ss}_{21}(-u)=\rho_2(u)=a_2(u)a_2(-u),\nonumber\\
{\rm crossing-unitarity}&:&R^{  ss}_{12}(u)^{t_1}R^{
ss}_{21}(-u-3)^{t_1}=\rho_2(u+\frac{3}{2}),
\end{eqnarray}
where ${\cal P}^{(s)}_{12}$ is the permutation operator among the spinorial representation space (c.f., ${\cal P}_{12}$ in (\ref{Permutation-1}) ).

Following the fusion procedure \cite{Kar79-1, Kar79-10, Kar79-2, Kar79-3, Kar79-4, Mez92-1, Mez92, Zho96}, we can construct another $R$-matrix $R^{sv}_{12}(u)$ defined in
$V^{(s)}\otimes V$. It is easily to check that
\bea  && R^{  ss}_{12}(-\frac{1}{2})=  P^{{
ss}(5) }_{12}\times S,\label{Int-Rs}
\eea
where $S$ is some non-degenerate constant matrix, and $P^{{  ss}(5) }_{12}$ is a 5-dimensional projector operator with the form
\bea
&& P^{{  ss}(5) }_{12}=\sum_{i=1}^5
|\tilde{\psi}_i\rangle\langle\tilde{\psi}_i|, \label{Project-v}
\eea
where the corresponding vectors are\footnote{We used a temporal notation $|ij\rangle^{(s)}=|i\rangle^{(s)}\otimes |j\rangle^{(s)}$. }
\bea
&&|\tilde{\psi}_1\rangle=\frac{1}{\sqrt{2}}(|12\rangle^{(s)}-|21\rangle^{(s)}),\qquad \qquad
|\tilde{\psi}_2\rangle=\frac{1}{\sqrt{2}}(|31\rangle^{(s)}-|13\rangle^{(s)}),\nonumber\\
&&|\tilde{\psi}_3\rangle=\frac{1}{{2}}(|14\rangle^{(s)}-|41\rangle^{(s)}+|23\rangle^{(s)}-|32\rangle^{(s)}),\quad
|\tilde{\psi}_4\rangle=\frac{1}{\sqrt{2}}(|24\rangle^{(s)}-|42\rangle^{(s)}),\nonumber\\
&&|\tilde{\psi}_5\rangle=\frac{1}{\sqrt{2}}(|34\rangle^{(s)}-|43\rangle^{(s)}).\nonumber \eea
Let $V^{(\langle ss\rangle)}$ denote the projected subspace of $V^{(s)}\otimes V^{(s)}$ by the projector $P^{{  ss}(5) }_{12}$. Namely, $V^{(\langle ss\rangle)}$
is a 5-dimensional subspace and spanned by $\{|\tilde\psi_i\rangle|i=1,\cdots,5\}$. Then we can construct a  fused $R_{1\,\langle 23\rangle }^{  sss}(u)$ matrix
\bea
R_{1\,\langle 23\rangle }^{  sss}(u)=[(u-\frac{1}{4})(u+\frac{3}{4})(u+\frac{7}{4})]^{-1}
P^{{  ss}(5) }_{23}R^{  ss}_{12}(u+\frac{1}{4})R^{  ss}_{13}(u-\frac{1}{4})P^{{  ss}(5) }_{23}.\label{Fused-R}
\eea
After taking the correspondence
\bea
|\tilde{\psi}_i\rangle \longrightarrow |i\rangle, \quad i=1,\cdots,5, \label{Identification}
\eea
one has the  identification: $V\equiv V^{(\langle ss\rangle)}$, which leads to  an $R$-matrix
$R^{sv}_{12}(u)$ defined in $V^{(s)}\otimes V$. The non-vanishing matrix elements of the resulting $R$-matrix are\footnote{Each matrix element of $R^{sv}_{12}(u)$,
as a function of $u$, is a polynomial with degree up to  one.}
\begin{eqnarray}
&&R^{  sv}(u)^{11}_{11}=R^{  sv}(u)^{12}_{12}=R^{  sv}(u)^{21}_{21}=R^{  sv}(u)^{24}_{24}=R^{  sv}(u)^{32}_{32}=R^{  sv}(u)^{35}_{35}\no\\[4pt]
&&\qquad =R^{  sv}(u)^{44}_{44}=R^{  sv}(u)^{45}_{45}=a_3(u)=u+\frac{5}{4},\nonumber\\[4pt]
&&R^{  sv}(u)^{14}_{14}=R^{  sv}(u)^{15}_{15}=R^{  sv}(u)^{22}_{22}=R^{  sv}(u)^{25}_{25}=R^{  sv}(u)^{31}_{31}=R^{  sv}(u)^{34}_{34}\no\\[4pt]
&&\qquad  =R^{  sv}(u)^{41}_{41}=R^{  sv}(u)^{42}_{42}=b_3(u)=u+\frac{1}{4},\nonumber\\[4pt]
&&R^{  sv}(u)^{14}_{41}=-R^{  sv}(u)^{15}_{42}=R^{  sv}(u)^{22}_{31}=-R^{  sv}(u)^{25}_{34}=R^{  sv}(u)^{31}_{22}=-R^{  sv}(u)^{34}_{25}\no\\[4pt]
&&\qquad =R^{  sv}(u)^{41}_{14}=-R^{  sv}(u)^{42}_{15}=-c_3(u)=-1,\nonumber\\[4pt]
&&R^{  sv}(u)^{13}_{13}=R^{  sv}(u)^{23}_{23}=R^{  sv}(u)^{33}_{33}=R^{  sv}(u)^{43}_{43}=e_3(u)=u+\frac{3}{4},\nonumber\\[4pt]
&&R^{  sv}(u)^{13}_{22}=-R^{  sv}(u)^{14}_{23}=R^{  sv}(u)^{13}_{31}=-R^{  sv}(u)^{15}_{33}=R^{  sv}(u)^{22}_{13}=-R^{  sv}(u)^{23}_{14}\no\\[4pt]
&&\qquad  =-R^{  sv}(u)^{23}_{41}=R^{  sv}(u)^{25}_{43}=R^{  sv}(u)^{31}_{13}=-R^{  sv}(u)^{33}_{15}=R^{  sv}(u)^{33}_{42}=-R^{  sv}(u)^{34}_{43}\nonumber\\[4pt]
&&\qquad  =-R^{  sv}(u)^{41}_{23}=R^{  sv}(u)^{43}_{25} =R^{  sv}(u)^{42}_{33}=-R^{  sv}(u)^{43}_{34}=-g_3(u)=-\frac{1}{\sqrt{2}}.
\label{RsV-elemensst}
\end{eqnarray}
It is easily to check that the fused $R_{12}^{  sv}$ matrix has the properties
\begin{eqnarray}
{\rm unitarity}&:&R^{  sv}_{1\,2}(u)R^{  vs}_{2\,1}(-u)=a_3(u)a_3(-u)\stackrel{{\rm def}}{=}\rho_3(u),\label{Uni-SV-1}\\
{\rm crossing-unitarity}&:&R^{  sv}_{1\,2}(u)^{t_1}R^{
vs}_{2\,1}(-u-3)^{t_1}=\rho_3(u+\frac{3}{2}).\label{Uni-SV-2}
\end{eqnarray}
Moreover, the fused $R_{1\,2}^{  sv}$ also
satisfy the Yang-Baxter equations
\bea R^{  sv}_{1\,2}(u_1-u_2)R^{  sv}_{1\,3}(u_1-u_3)R^{
vv}_{2\,3}(u_2-u_3) =R^{  vv}_{2\,3}(u_2-u_3)R^{
sv}_{1\,3}(u_1-u_3)R^{  sv}_{1\,2}(u_1-u_2),\label{QYB-2}\eea
and \bea
R^{  ss}_{1\,2}(u_1-u_2)R^{  sv}_{1\,3}(u_1-u_3)R^{
sv}_{2\,3}(u_2-u_3) =R^{  sv}_{2\,3}(u_2-u_3)R^{
sv}_{1\,3}(u_1-u_3)R^{  ss}_{1\,2}(u_1-u_2).\label{QYB-3}\eea
Similarly, one can reconstruct the $R$-matrix $R^{vv}_{12}(u)$ from the fused one $R^{sv}_{12}(u)$
\bea
R^{vv}_{1\,3}(u)\stackrel{(\ref{Identification})}{\equiv}R_{\langle 1\,2\rangle\, 3}^{  ssv}(u)=
P^{{  ss}(5) }_{12}R^{  sv}_{2\,3}(u+\frac{1}{4})R^{  sv}_{1\,3}(u-\frac{1}{4})P^{{  ss}(5) }_{12}.
\eea

We have  checked that  the $R$-matrices $R^{sv}_{12}(u)$ and $R^{vv}_{12}(u)$ enjoy the properties:
\bea &&
R^{vv}_{12}(-\frac{3}{2})=  P^{{  vv}(1) }_{12}\times S_1,\label{Int-R1}\\[4pt]
&& R^{vv}_{12}(-1)=  P_{12}\times
S_2,\label{Int-R3} \\[4pt]
&& R^{ vv}_{12}(-1)R^{  vv}_{13}(-2)R^{  vv}_{23}(-1)=
P_{123}\times
S_3,\label{Int-R4} \\[4pt]
&& R^{ vv}_{12}(-1)R^{  vv}_{13}(-2)R^{  vv}_{14}(-3)R^{
vv}_{23}(-1)R^{  vv}_{24}(-2)R^{  vv}_{34}(-1)= P_{1234}\times S_4,\label{Int-R5} \\[4pt]
&& R^{ sv}_{12}(-\frac{5}{4})=  P^{{  sv}(4) }_{12}\times S_5,\label{Int-R2}
\eea where $ P^{{
vv}(1) }_{12}$, $ P_{12}$, $P_{123}$, $P_{1234}$ and $P^{{  sv}(4) }_{12}$ are the projectors given by (\ref{a1})-(\ref{a5}) in Appendix  A and
$S_i ( i=1,2,\dots,5)$ are some irrelevant constant matrices. With the help of the above projectors and using the similar fusion procedure
\cite{Kar79-1, Kar79-10, Kar79-2, Kar79-3, Kar79-4, Mez92-1, Mez92, Zho96},
we can construct the fused $R$-matrices
\bea
&&\hspace{-1.42truecm}R_{\bar{1}\, 3}^{  \bar{v}v}(u)\stackrel{{\rm def}}{=} R_{\langle 12\rangle 3}^{  \bar{v}v}(u)
=\tilde{\rho}_0^{-1}(u+\frac{1}{2})P_{21}R^{  vv}_{13}(u+\frac{1}{2})R^{  vv}_{23}(u-\frac{1}{2})P_{21},\label{Fused-R-1} \\[4pt]
&&\hspace{-1.42truecm}R_{ \tilde{1}\, 3}^{  \tilde{v}v}(u)\stackrel{{\rm def}}{=} R_{\langle 123\rangle 4}^{  \tilde{v}v}(u)
=[\tilde{\rho}_0(u+1)\tilde{\rho}_0(u)]^{-1}
P_{321}R^{  vv}_{14}(u\hspace{-0.12truecm}+\hspace{-0.12truecm}1)R^{vv}_{24}(u)R^{  vv}_{34}(u\hspace{-0.12truecm}-\hspace{-0.12truecm}1)P_{321},\label{Fused-R-2} \\[4pt]
&&\hspace{-1.42truecm}R_{\langle 1234\rangle
5}(u)=\tilde{\rho}_1^{-1}(u) P_{4321}R^{vv}_{15}(u)R^{
vv}_{25}(u-1)R^{
vv}_{35}(u-2)R^{vv}_{45}(u\hspace{-0.12truecm}-3)P_{4321},
\label{Fused-R-3} \eea where \bea
\tilde{\rho}_0(u)=(u-1)(u+\frac{3}{2}),\quad
\tilde{\rho}_1(u)=\tilde{\rho}_0(u)\tilde{\rho}_0(u-1)\tilde{\rho}_{0}(u-2).
\eea Here we have used $\bar{V}$ (resp. $\tilde{V}$) to denote the
projected subspace in $V\otimes V$  by  $P_{21}$ (resp. the
projected subspace  in $V\otimes V\otimes V$ by $P _{321}$ ) and
adopted the convention: $\bar{1}\equiv \langle 12\rangle$ and
$\tilde{1}\equiv \langle 123\rangle$.

Some remarks are in order. It is shown that each matrix elements of the above fused $R$-matrices, as a function of $u$,  is a polynomial with degree up to two. Due to the fact that the $16$-dimensional  projected subspace in $V\otimes V\otimes V\otimes V$ by the projector $P^{{  vv} }_{4321}$ is equivalent to the tensor space $V^{(s)}\otimes V^{(s)}$ according to the correspondence (\ref{Identification-3}) below, we have the  equivalence
\bea
&&R_{\langle 1234\rangle 5}(u) \equiv S_{12}\,R^{  sv}_{1\,5}(u-\frac{1}{4})\,R^{
sv}_{2\,5}(u-\frac{11}{4})\,S_{12}^{-1},\label{Spinor}
\eea
where the constant gauge transformation matrix $S_{12}$ is given by (\ref{RsV-element}) below.
Moreover, we have the identity
\bea
&&P^{{  vv}(1) }_{21}R^{  vv}_{13}(u)R^{  vv}_{23}(u-\frac{3}{2})P^{{  vv}(1) }_{21}=a_1(u)e_1(u-\frac{3}{2})P^{{  vv}(1) }_{21}\times {\rm id}.\label{Quantum-det}
\eea

\subsection{Operator product identities}

Besides the transfer matrix $t^{(p)}_1(u)$ given by (\ref{1117-1}), let us introduce $3$  fused transfer matrices:
\bea
\bar{t}^{\,(p)}_m(u)=tr_{\langle 12\cdots m\rangle} {\bar T}_{\langle 12\cdots m\rangle}^{  v}(u), \quad m=2,3,4,\label{Fused-transfer-1}
\eea
where ${\bar T}_{\langle 12\cdots m\rangle}^{  v}(u)$ are the fused monodromy matrices
\bea
\bar T_{\langle 12\cdots m\rangle}^{  v}(u)=P_{m\cdots 21}\,T_1^{
v}(u)T_2^{  v}(u-1) T_3^{  v}(u-2) \cdots T_m^{  v}(u-m+1)\,P_{m\cdots 21}, \label{Fused-M-1}
\eea
and the projectors $\{P_{m\cdots 21}|m=2,3,4\}$ are given by (\ref{a1})-(\ref{a4}) below. Moreover, in order to
have  closed ( or enough) operator product identities, we need to introduce an extra transfer matrix
\bea
t^{(p)}_s(u)=tr_{0} T_{0}^{  s}(u), \quad \quad T_{0}^{  s}(u)=R^{    sv}_{01}(u-\theta_1)R^{   sv}_{02}(u-\theta_2)\cdots R^{
   sv}_{0N}(u-\theta_N),\label{Transfer-S}
\eea
where the spinoral $R$-matrix $R^{sv}_{12}(u)$ is given by (\ref{RsV-elemensst}).
It is easily to show that all the transfer matrices constitute a commutative family, namely,
\bea
[t^{(p)}_s(u),\,t^{(p)}_s(v)]=[t^{(p)}_s(u),\,{\bar{t}}^{\,(p)}_m(v)]=[{\bar{t}}^{\,(p)}_m(u),\,{\bar{t}}^{\,(p)}_n(v)]=0,\quad m,n=1,2,\cdots,4.
\eea
We have used the convention: ${\bar{t}}^{\,(p)}_1(u)=t^{(p)}_1(u)=t^{(p)}(u)$.
Direct calculation shows that
\bea
&&P^{{  vv}(1)}_{21}T_1^{  v}(u)T_2^{  v}(u-\frac{3}{{2}})P^{{  vv}(1)}_{21}=\prod_{i=1}^N
a_1(u-\theta_i)e_1(u-\theta_i-\frac{3}{{2}})\,P^{{  vv}(1)}_{21}\times {\rm id}, \label{Relation-1} \\
&&\bar T_{\langle 12\rangle }^{  v}(u)=\prod_{i=1}^N\tilde{\rho}_0(u-\theta_i)\,{T}_{\bar 1 }^{  \bar{v}}(u-\frac{1}{{2}}), \label{Relation-2} \\
&&\bar T_{\langle 123\rangle }^{  v}(u) =\prod_{i=1}^N
\tilde{\rho}_0(u-\theta_i)\tilde{\rho}_0(u-\theta_i-1)\,{T}_{\tilde
1 }^{ \tilde {v}}(u-1),\label{Relation-3} \eea where we have
introduced some normalized monodromy matrices:
\begin{eqnarray}
 &&T_{\bar{0}}^{  \bar{v}}(u)=R^{    \bar{v}v}_{\bar{0}1}(u-\theta_1)R^{   \bar{v}v}_{\bar{0}2}(u-\theta_2)\cdots R^{
   \bar{v}v}_{\bar{0}N}(u-\theta_N), \label{T4-0} \\
 &&T_{\tilde{0}}^{  \tilde{v}}(u)=R^{    \tilde{v}v}_{\tilde{0}1}(u-\theta_1)R^{   \tilde{v}v}_{\tilde{0}2}(u-\theta_2)\cdots R^{
   \tilde{v}v}_{\tilde{0}N}(u-\theta_N).  \label{T4}
\end{eqnarray}
It is remarked that the quantum spaces  of the above monodromy matrices are the same (i.e., $V^{\otimes N}$) and that the corresponding auxiliary spaces are $\bar{V}$ and $\tilde{V}$ with dimensions $11$ and $15$. Then the associated  transfer matrices are given by
\bea
  &&t^{(p)}_2(u)=tr_{\bar{0}} T_{\bar{0}}^{  \bar{v}}(u),\quad t^{(p)}_3(u)=tr_{\tilde{0}} T_{\tilde{0}}^{  \tilde{v}}(u).\label{Fused-transfer-matrix-periodic}
\eea
The equivalence (\ref{Spinor}) and the relations (\ref{Relation-1})-(\ref{Relation-3}) imply that
\begin{eqnarray}
&&\bar t^{\;(p)}_2(u)= \prod_{i=1}^N \tilde{\rho}_0(u-\theta_i)\,  t^{(p)}_2(u-\frac12), \no \\
&&\bar t^{\;(p)}_3(u)= \prod_{i=1}^N \tilde{\rho}_0(u-\theta_i) \tilde{\rho}_0(u-\theta_i-1)\, t^{(p)}_3(u-1),\no \\
&& \bar t^{\;(p)}_4(u)= \prod_{i=1}^N \tilde{\rho}_1(u-\theta_i)\, t^p_s(u-\frac14)\, t^{(p)}_s(u-\frac{11}{4}).\label{bet1}
\end{eqnarray}
Fowllowing the method developed in \cite{Cao14JHEP143},  we obtain the identities
\bea &&T_1^{  v}(\theta_j)\,T_2^{v}(\theta_j-\frac{3}{{2}})=
P^{{  vv}(1) }_{21}\,T_1^{v}(\theta_j)\,T_2^{  v}(\theta_j-\frac{3}{{2}}), \no \\[6pt]
&&T_1^{  v}(\theta_j)\,T_2^{  v}(\theta_j-1)=P_{21}\,T_1^{v}(\theta_j)\,T_2^{  v}(\theta_j-1),\no \\[6pt]
&&T_1^{  v}(\theta_j)\,\bar T_{\langle23\rangle}^{  v}(\theta_j-1)=
P_{321}\,T_1^{v}(\theta_j)\,\bar T_{\langle23\rangle}^{  v}(\theta_j-1),\no \\[6pt]
&&T_1^{  v}(\theta_j)\,\bar T_{\langle 234\rangle}^{  v}(\theta_j-1)=
P_{4321}\,T_1^{v}(\theta_j)\,\bar T_{\langle 234\rangle}^{  v}(\theta_j-1),\no \\[6pt]
&&T_2^{  v}(\theta_j)\,T_{\bar 1}^{\bar{v}}(\theta_j-1)=P^{{  \bar{v}v}(5) }_{\bar{1}2}\,
T_2^{v}(\theta_j)\,T_{\bar 1}^{\bar{v}}(\theta_j-1),\no \\[6pt]
&&T_2^{  v}(\theta_j)\,T_{\tilde 1}^{\tilde{v}}(\theta_j-\frac{1}{{2}})=
P^{{  \tilde{v}v}(11)}_{\tilde{1}2}\,T_2^{  v}(\theta_j)\,
T_{\tilde 1}^{\tilde{v}}(\theta_j-\frac{1}{{2}}),\no \\[6pt]
&&T_2^{  v}(\theta_j)\,T_{ 1}^{s}(\theta_j-\frac{5}{{4}})=
P^{{  sv}(4) }_{{1}2}\,T_2^{v}(\theta_j)\,T_{ 1}^{  s}(\theta_j-\frac{5}{{4}}),\label{fui-7}
\eea
where the explicit expressions of the projectors $P^{{  vv}(1) }_{21}$, $P_{21}$, $P_{321}$, $P_{4321}$, $P^{{  \bar{v}v}(5) }_{\bar{1}2}$,
$P^{{  \tilde{v}v}(11)}_{\tilde{1}2}$ and $P^{{  sv}(4) }_{{1}\,2}$ are given in Appendix A. With the help of the correspondences (\ref{Spinor}),
(\ref{R5-5}), (\ref{R11-R11}) and (\ref{R4-R4}), we have that the transfer matrices
satisfy the operator product identities:\footnote{Because for generic choices of inhomogeneous parameters $\{\theta_j\}$ these identities involve
the different points values of the transfer matrices (for $t^{(p)}(u)$ at $\theta_j, \theta_j-\frac{1}{2}, \theta_j-1, \theta_j-\frac{3}{2};$ for $t_2^{(p)}(u)$ at
$\theta_j, \theta_j-\frac{1}{2}, \theta_j-1, \theta_j-\frac{3}{2};$ for $t_3^{(p)}(u)$ at $\theta_j-\frac{1}{2}, \theta_j-1, \theta_j-2;$ for $t_s^{(p)}(u)$ at
$\theta_j-\frac{1}{4}, \theta_j-\frac{5}{4}, \theta_j-\frac{11}{4}$), the identities (\ref{Op-Product-Periodic-1})-(\ref{futp-7}) are independent.}
\bea && t^{(p)}(\theta_j)\,t^{(p)}(\theta_j-\frac{3}{{2}})=\prod_{i=1}^N
a_1(\theta_j-\theta_i)e_1(\theta_j-\theta_i-\frac{3}{{2}})\times {\rm id},\label{Op-Product-Periodic-1}  \\
&& t^{(p)}(\theta_j)\,t^{(p)}(\theta_j-1)= \prod_{i=1}^N
\tilde{\rho}_0(\theta_j-\theta_i)\,t_2^{(p)}(\theta_j-\frac{1}{{2}}),\label{Op-Product-Periodic-2} \\
&& t^{(p)}(\theta_j)\,t_2^{(p)}(\theta_j-\frac{3}{{2}})=
\prod_{i=1}^N
\tilde{\rho}_0(\theta_j-\theta_i)\,t_3^{(p)}(\theta_j-1),\label{Op-Product-Periodic-3}  \\
&& t^{(p)}(\theta_j)\,t_3^{(p)}(\theta_j-2)= \prod_{i=1}^N
\tilde{\rho}_0(\theta_j-\theta_i)\,t^{(p)}_s(\theta_j-\frac{1}{{4}})\,t^{(p)}_{s}(\theta_j-\frac{11}{{4}}),\label{Op-Product-Periodic-4}  \\
&& t^{(p)}(\theta_j)\,t^{(p)}_2(\theta_j-1)=\prod_{i=1}^N
\tilde{\rho}_0(\theta_j-\theta_i)\,t^{(p)}(\theta_j-\frac{1}{{2}}),\label{Op-Product-Periodic-5}  \\
&& t^{ (p)}(\theta_j)t_3^{(p)}(\theta_j-\frac{1}{{2}})=\prod_{i=1}^N
\tilde{\rho}_0(\theta_j-\theta_i)\,t_2^{(p)}(\theta_j),\label{Op-Product-Periodic-6} \\
&& t^{(p)}(\theta_j)\,t^{(p)}_{s}(\theta_j-\frac{5}{{4}})=\prod_{i=1}^N
\tilde{\rho}_0(\theta_j-\theta_i)\,t^{(p)}_{s}(\theta_j-\frac{1}{{4}}).\label{futp-7}
\eea
Now, we consider the asymptotic behaviors of the fused transfer matrices. Direct calculation shows
\bea && t^{ (p)}(u)|_{u\rightarrow \pm\infty}= 5u^{2N}\times {\rm  id} +\cdots,\no \\[4pt]
&& t^{(p)}_2(u)|_{u\rightarrow \pm\infty}=
11u^{2N}\times {\rm  id} +\cdots,\no \\[4pt]
&& t^{(p)}_3(u)|_{u\rightarrow \pm\infty}=
15u^{2N}\times {\rm   id} +\cdots,\no \\[4pt]
&& t^{(p)}_{s}(u)|_{u\rightarrow \pm\infty}=
4u^{N}\times { \rm  id} +\cdots.\label{fuwwwtpl-7} \eea

Let us denote the eigenvalues of the transfer matrices $t^{(p)}(u)$, $t^{(p)}_2(u)$ $t^{(p)}_3(u)$ and $t^{(p)}_s(u)$ as
$\Lambda^{(p)}(u)$, $\Lambda^{(p)}_2(u)$, $\Lambda^{(p)}_3(u)$ and $\Lambda^{(p)}_s(u)$, respectively.
From the operator product identities (\ref{Op-Product-Periodic-1})-(\ref{futp-7}), we have the functional relations
among the eigenvalues\footnote{It is remarked that only (\ref{Eigen-function-relation-1}) and (\ref{Eigen-function-relation-7}) were  used to obtain  $\Lambda^{(p)}(u)$ and $\Lambda^{(p)}_s(u)$ for the closed $B_n$ chain \cite{NYReshetikhin1,NYReshetikhin2}.
Since that $\Lambda^{(p)}(u)$ (resp. $\Lambda^{(p)}_s(u)$) is a polynomial of $u$ with degree $2N$ (resp. degree $N$),  one need $3N+2$ conditions to
determine them completely while (\ref{Eigen-function-relation-1}) and (\ref{Eigen-function-relation-7}), together with the asymptotic behaviors, only give $2N+2$ conditions. Therefore, in order to close the functional relations, (\ref{Eigen-function-relation-1})-(\ref{Eigen-function-relation-7}) are necessary.}:
\bea && \Lambda^{ (p)}(\theta_j)\,\Lambda^{(p)}(\theta_j-\frac{3}{{2}})=\prod_{i=1}^N
a_1(\theta_j-\theta_i)\,e_1(\theta_j-\theta_i-\frac{3}{{2}}),\label{Eigen-function-relation-1}  \\
&& \Lambda^{ (p)}(\theta_j)\,\Lambda^{ (p)}(\theta_j-1)=
\prod_{i=1}^N
\tilde{\rho}_0(\theta_j-\theta_i)\,\Lambda^{(p)}_2(\theta_j-\frac{1}{{2}}),\label{Eigen-function-relation-2}  \\
&&\Lambda^{(p)}(\theta_j)\,\Lambda^{(p)}_2(\theta_j-\frac{3}{{2}})= \prod_{i=1}^N
\tilde{\rho}_0(\theta_j-\theta_i)\,\Lambda^{(p)}_3(\theta_j-1), \label{Eigen-function-relation-3} \\
&& \Lambda^{(p)}(\theta_j)\,\Lambda^{(p)}_3(\theta_j-2)=
\prod_{i=1}^N
\tilde{\rho}_0(\theta_j-\theta_i)\,\Lambda^{(p)}_{  s}(\theta_j-\frac{1}{{4}})\,
\Lambda^{(p)}_{s}(\theta_j-\frac{11}{{4}}), \label{Eigen-function-relation-4}  \\
&& \Lambda^{(p)}(\theta_j)\,\Lambda^{(p)}_2(\theta_j-1)=\prod_{i=1}^N
\tilde{\rho}_0(\theta_j-\theta_i)\,\Lambda^{(p)}(\theta_j-\frac{1}{{2}}),\label{Eigen-function-relation-5}  \\
&& \Lambda^{(p)}(\theta_j)\,\Lambda^{(p)}_3(\theta_j-\frac{1}{{2}})=\prod_{i=1}^N
\tilde{\rho}_0(\theta_j-\theta_i)\,\Lambda^{(p)}_2(\theta_j),\label{Eigen-function-relation-6}  \\
&& \Lambda^{(p)}(\theta_j)\Lambda^{(p)}_{s}(\theta_j-\frac{5}{{4}})=\prod_{i=1}^N
\tilde{\rho}_0(\theta_j-\theta_i)\,\Lambda^{(p)}_{s}(\theta_j-\frac{1}{{4}}).\label{Eigen-function-relation-7}
\eea
The  asymptotic behaviors (\ref{fuwwwtpl-7}) of the fused transfer matrices lead to the corresponding
asymptotic behaviors of their eigenvalues:
\bea && \Lambda^{ (p)}(u)|_{u\rightarrow \pm\infty}= 5u^{2N} +\cdots,\label{Eigen-function-relation-8} \\[4pt]
&& \Lambda^{(p)}_2(u)|_{u\rightarrow \pm\infty}=
11u^{2N}+\cdots,\label{Eigen-function-relation-9} \\[4pt]
&& \Lambda^{(p)}_3(u)|_{u\rightarrow \pm\infty}=
15u^{2N} +\cdots,\label{Eigen-function-relation-10} \\[4pt]
&& \Lambda^{(p)}_{s}(u)|_{u\rightarrow \pm\infty}=
4u^{N} +\cdots.\label{Eigen-function-relation-11} \eea

From the definitions (\ref{1117-1}), (\ref{Transfer-S}) and (\ref{Fused-transfer-matrix-periodic}), we know that the eigenvalues
$\Lambda^{(p)}(u)$, $\Lambda^{(p)}_2(u)$ and $\Lambda^{ (p)}_3(u)$ are polynomials of $u$ with degree $2N$, while $\Lambda^{(p)}_{  s}(u)$ is a
polynomial of $u$ with degree $N$. Hence the functional relations (\ref{Eigen-function-relation-1})-(\ref{Eigen-function-relation-11}) could completely determine
the eigenvalues, which allows us to express them in terms of some homogeneous $T-Q$ relations in the next subsection.

\subsection{$T-Q$ relations}

Let us introduce some  functions:
\bea &&Z^{(p)}_1(u)=\prod_{j=1}^N
a_1(u-\theta_j)\,\frac{Q_{p}^{(1)}(u-1)}{Q_{p}^{(1)}(u)},\no\\[4pt]
&&Z^{(p)}_2(u)=\prod_{j=1}^Nb_1(u-\theta_j)\,\frac{Q_{p}^{(1)}(u+1)Q_{p}^{(2)}(u-1)}{Q_{p}^{(1)}(u)Q_{p}^{(2)}(u)},\no\\[4pt]
&&Z^{(p)}_3(u)=\prod_{j=1}^N
b_1(u-\theta_j)\,\frac{Q_{p}^{(2)}(u-1)Q_{p}^{(2)}(u+\frac{1}{2})}{Q_{p}^{(2)}(u)Q_{p}^{(2)}(u-\frac{1}{2})},\no\\[4pt]
&&Z^{(p)}_4(u)=\prod_{j=1}^N
b_1(u-\theta_j)\,\frac{Q_{p}^{(1)}(u-\frac{1}{2})Q_{p}^{(2)}(u+\frac{1}{2})}{Q_{p}^{(1)}(u+\frac{1}{2})Q_{p}^{(2)}(u-\frac{1}{2})},\no\\
&&Z^{(p)}_5(u)=\prod_{j=1}^N e_1(u-\theta_j)\,
\frac{Q_{p}^{(1)}(u+\frac{3}{2})}{Q_{p}^{(1)}(u+\frac{1}{2})},\no\\[4pt]
&&Q_{p}^{(m)}(u)=\prod_{k=1}^{L_m}(u-\mu_k^{(m)}+\frac{m}{2}), \quad m=1,2.\eea
The functional relations (\ref{Eigen-function-relation-1})-(\ref{Eigen-function-relation-11}) enable us to parameterize the eigenvalues of
the transfer matrices in terms of the $T-Q$ relations as follows:
\bea &&\Lambda^{(p)}(u)=Z^{(p)}_1(u)+
Z^{(p)}_2(u)+Z^{(p)}_3(u)+Z^{(p)}_4(u)+Z^{(p)}_5(u), \label{T-Q-Hom-1} \\[4pt]
&&\Lambda^{(p)}_2(u)=\prod_{i=1}^N
\tilde{\rho}_0^{-1}(u-\theta_i+\frac12)\,\{
Z^{(p)}_1(u+\frac12)\,[
Z^{(p)}_2(u-\frac12)+Z^{(p)}_3(u-\frac12)+Z^{(p)}_4(u-\frac12)\no\\[4pt]
&&\hspace{10mm}+Z^{(p)}_5(u-\frac12)]
+Z^{(p)}_2(u+\frac12)\,[Z^{(p)}_3(u-\frac12)+Z^{(p)}_4(u-\frac12)+Z^{(p)}_5(u-\frac12)]\no\\[4pt]
&&\hspace{10mm}+Z^{(p)}_3(u+\frac12)[ Z^{(p)}_3(u-\frac12) +Z^{(p)}_4(u-\frac12)+Z^{(p)}_5(u-\frac12)]\no\\[4pt]
&&\hspace{10mm}+Z^{(p)}_4(u+\frac12)\,Z^{(p)}_5(u-\frac12)\},
\label{T-Q-Hom-2} \\[4pt]
&&\Lambda^{(p)}_3(u)=\prod_{i=1}^N [\tilde{\rho}_0(u-\theta_i+1)
\tilde{\rho}_0(u-\theta_i)]^{-1} \{Z^{(p)}_1(u+1)\,Z^{(p)}_2(u)\,[
Z^{(p)}_3(u-1)+Z^{(p)}_4(u-1)\no\\[4pt]
&&\hspace{10mm}+Z^{(p)}_5(u-1)]+Z^{(p)}_1(u+1)\,Z^{(p)}_3(u)\,[
Z^{(p)}_3(u-1)+Z^{(p)}_4(u-1)+Z^{(p)}_5(u-1)]\no\\[4pt]
&&\hspace{10mm}+Z^{(p)}_2(u+1)\,Z^{(p)}_3(u)\,[
Z^{(p)}_3(u-1)+Z^{(p)}_4(u-1)+Z^{(p)}_5(u-1)]\no\\[4pt]
&&\hspace{10mm}+Z^{(p)}_3(u+1)\,Z^{(p)}_3(u)\,[ Z^{(p)}_3(u-1)+Z^{(p)}_4(u-1)+Z^{(p)}_5(u-1)]\no\\[4pt]
&&\hspace{10mm}+(Z^{(p)}_1(u+1)+Z^{(p)}_2(u+1)+Z^{(p)}_3(u+1))\,Z^{(p)}_4(u)\,Z^{(p)}_5(u-1)\}, \label{T-Q-Hom-3} \\[4pt]
&&\Lambda^{(p)}_{  s}(u)= \prod_{j=1}^N[(u+\frac{7}{4}-\theta_j)
b_1(u-\frac{3}{4}-\theta_j)]^{-1}\,
\frac{Q_{p}^{(2)}(u-\frac{5}{4})}{Q_{p}^{(2)}(u-\frac{7}{4})}\no\\[4pt]
&&\hspace{12mm}\times \{Z^{(p)}_1(u+\frac{1}{4})Z^{(p)}_2(u-\frac{3}{4})+Z^{(p)}_1(u+\frac{1}{4})Z^{(p)}_3(u-\frac{3}{4})\no\\[4pt]
&&\hspace{12mm}+Z^{(p)}_2(u+\frac{1}{4})\,Z^{(p)}_3(u-\frac{3}{4})+Z^{(p)}_3(u+\frac{1}{4})\,Z^{(p)}_3(u-\frac{3}{4})\}
\label{ep-1} \no\\[4pt]
 &&\hspace{12mm}=
(\prod_{j=1}^N(u-\frac{1}{4}-\theta_j)
b_1(u+\frac{3}{4}-\theta_j))^{-1}
\frac{Q^{(2)}_p(u+\frac{3}{4})}{Q^{(2)}_p(u+\frac{5}{4})}\no\\[4pt]
&&\hspace{12mm} \times
\{Z^{(p)}_3(u+\frac{3}{4})Z^{(p)}_3(u-\frac{1}{4})+Z^{(p)}_3(u+\frac{3}{4})Z^{(p)}_4(u-\frac{1}{4})\no\\[4pt]
&&\hspace{12mm}+Z^{(p)}_3(u+\frac{3}{4})Z^{(p)}_5(u-\frac{1}{4})+Z^{(p)}_4(u+\frac{3}{4})Z^{(p)}_5(u-\frac{1}{4})\}.\label{ep-2}
\eea The regularity of the eigenvalue $\Lambda^{(p)}(u)$ given by
(\ref{T-Q-Hom-1}) leads to that the parameters $\{\mu^{(m)}_k\}$
should satisfy the Bethe Ansatz equations (BAEs): \bea &&
\frac{Q_{p}^{(1)}(\mu_k^{(1)}+\frac{1}{2})Q_{p}^{(2)}(\mu_k^{(1)}-\frac{3}{2})}{Q_{p}^{(1)}(\mu_k^{(1)}-\frac{3}{2})Q_{p}^{(2)}(\mu_k^{(1)}-\frac{1}{2})}
=-\prod_{j=1}^N \frac{\mu_k^{(1)}+\frac{1}{2}-\theta_j
}{\mu_k^{(1)}-\frac{1}{2}-\theta_j}, \quad k=1,\cdots, L_1, \label{BAEs-1} \\[8pt]
&&\frac{Q_{p}^{(1)}(\mu_l^{(2)})Q_{p}^{(2)}(\mu_l^{(2)}-\frac{3}{2})}{Q_{p}^{(1)}(\mu_l^{(2)}-1)Q_{p}^{(2)}(\mu_l^{(2)}-\frac{1}{2})}
=-1, \quad l=1,\cdots, L_2.\label{BAEs-2} \eea
We have verified that the above BAEs indeed guarantee all the $T-Q$ relations (\ref{T-Q-Hom-1})-(\ref{ep-1}) are  polynomials of $u$ with the required degrees. Moreover,  these $T-Q$ relations also  satisfy the function relations (\ref{Eigen-function-relation-1})-(\ref{Eigen-function-relation-7}) and the asymptotic behaviors (\ref{Eigen-function-relation-8})-(\ref{Eigen-function-relation-11}).
Therefore, we conclude that $\Lambda^{(p)}(u)$, $\Lambda^{(p)}_2(u)$, $\Lambda^{(p)}_3(u)$ and $\Lambda^{(p)}_s(u)$ given by (\ref{T-Q-Hom-1})-(\ref{ep-1}) are
indeed the eigenvalues of  the transfer matrices $t^{(p)}(u)$, $t^{(p)}_2(u)$, $t^{(p)}_3(u)$ and $t^{(p)}_s(u)$ provided that the $L_1+L_2$ parameters $\{\mu^{(m)}_k\}$ satisfy the associated BAEs  (\ref{BAEs-1})-(\ref{BAEs-2}).
It is remarked that the $T-Q$ relations (\ref{T-Q-Hom-1}) and the
associated BAEs (\ref{BAEs-1})-(\ref{BAEs-2}) (after taking the homogeneous limit $\{\theta_j\to0|j=1,2,\cdots,N\}$) coincide with those
obtained previously via conventional Bethe Ansatz
methods\cite{NYReshetikhin1,NYReshetikhin2,Bn}. Comparing with the results of the structure for Bethe states \cite{Bn}, we conclude that the two no-negative integers $L_1$ and $L_2$ take values: $0\leq L_1\leq 2N,\quad 0\leq L_2\leq 2L_1$, which gives rise to the complete spectrum of the transfer matrix (\ref{1117-1}).

\section{Off-diagonal open boundary case}
\setcounter{equation}{0}

\subsection{Open chain}

Integrable open chain can be constructed as follows \cite{Alc87,Skl88}.
Let us introduce a pair of $K$-matrices $K^{v-}(u)$ and $K^{v+}(u)$. The former satisfies the reflection equation (RE)
\begin{equation}
 R^{   vv}_{12}(u-v){K^{  v-}_{  1}}(u)R^{   vv}_{21}(u+v) {K^{   v-}_{2}}(v)=
 {K^{   v-}_{2}}(v)R^{   vv}_{12}(u+v){K^{   v-}_{1}}(u)R^{   vv}_{21}(u-v),
 \label{r1}
 \end{equation}
and the latter  satisfies the dual RE
\begin{eqnarray}
 &&R^{   vv}_{12}(-u+v){K^{   v+}_{1}}(u)R^{  vv}_{21}
 (-u-v-3){K^{   v+}_{2}}(v)\nonumber\\[4pt]
&&\qquad\qquad\quad\quad={K^{   v+}_{2}}(v)R^{   vv}_{12}(-u-v-3) {K^{
v+}_{1}}(u)R^{  vv}_{21}(-u+v).
 \label{r2}
 \end{eqnarray}
For open spin chains, instead of the
``row-to-row" monodromy matrix $T^{v}_0(u)$ (\ref{Mon-1}), one needs to
consider  the ``double-row" monodromy matrix  as follows. Let us introduce another ``row-to-row" monodromy matrix
\begin{eqnarray}
\hat{T}_0^{  v} (u)=R_{N0}^{  vv}(u+\theta_N)\cdots R_{20}^{  vv}(u+\theta_{2}) R_{10}^{  vv}(u+\theta_1),\label{Tt11}
\end{eqnarray}
which satisfies the Yang-Baxter relation
\begin{eqnarray}
R_{ 12}^{  vv} (u-v) \hat T_{1}^{  v}(u) \hat T_2^{  v}(v)=\hat  T_2^{  v}(v) \hat T_{ 1}^{  v}(u) R_{12}^{  vv} (u-v).\label{haishi0}
\end{eqnarray}
The transfer matrix $t(u)$ is defined as
\begin{equation}
t(u)\stackrel{{\rm def}}{=}t_1(u)= tr_0 \{K_0^{  v +}(u)T_0^{  v} (u) K^{  v -}_0(u)\hat{T}^{  v}_0 (u)\}. \label{trweweu}
\end{equation}
From the Yang-Baxter relation, reflection equation and its dual, one can
prove that the transfer matrices with different spectral parameters
commute with each other, $[t(u), t(v)]=0$. Therefore, $t(u)$ serves
as the generating function of all the conserved quantities of the
system.  The Hamiltonian of the open chain can be obtained by taking the
derivative of the logarithm of the transfer matrix
\begin{eqnarray}
H&=&\frac{\partial \ln t(u)}{\partial
u}|_{u=0,\{\theta_j\}=0} \nonumber \\[8pt]
&=& \sum^{N-1}_{k=1}H_{k k+1}+\frac{1}{2\xi}{K^{v-}_1}'(0)+\frac{
tr_0 \{K^{v+}_0(0)H_{N0}\}}{tr_0 K^{v+}_0(0)}+{\rm constant},
\label{hh}
\end{eqnarray}
with $K^{v-}(0)=\xi\times {\rm id}.$

In this paper, we consider an open chain associated with the
off-diagonal $K$-matrix  $K^{v-}(u)$ given by \bea
 K^{  v-}(u)=\left(\begin{array}{ccccc}K^{  v}_{11}(u)&0&K^{  v}_{13}(u)&0&K^{  v}_{15}(u)\\[6pt]
0&K^{  v}_{22}(u)&0&0&0\\[6pt]
K^{  v}_{31}(u)&0&K^{  v}_{33}(u)&0&K^{  v}_{35}(u)\\[6pt]
0&0&0&K^{  v}_{44}(u)&0\\[6pt]
K^{  v}_{51}(u)&0&K^{  v}_{53}(u)&0&K^{  v}_{55}(u)
\end{array}\right),\label{K-matrix-VV} \eea
where the non-vanishing matrix elements are
\bea
&&K^{  v}_{11}(u)=-1-c_1c_2+4u,\quad  K^{  v}_{13}(u)=4{\sqrt{2}}c_1{u}, \quad K^{  v}_{15}(u)=-4c_1^2u,\no\\[4pt]
&& K^{  v}_{22}(u)=-(1+c_1c_2)(4u+1),\quad K^{  v}_{31}(u)=4{\sqrt{2}}c_2{u}, \quad K^{  v}_{35}(u)=4{\sqrt{2}}c_1{u}, \no\\[4pt]
&&  K^{  v}_{33}(u)=-1-4u+c_1c_2(4u-1), \quad  K^{  v}_{44}(u)=-(1+c_1c_2)(4u+1),\no\\[4pt]
  &&K^{  v}_{51}(u)=-4c_2^2u,\quad K^{  v}_{53}(u)=4{\sqrt{2}}c_2{u}, \quad  K^{  v}_{55}(u)=-1-c_1c_2+4u.  \eea
Here $c_1$ and $c_2 $ are arbitrary boundary parameters.  The dual reflection matrix $K^{v+}(u)$ is also an off-diagonal one and
given by
\begin{equation}
K^{   v+}(u)=K^{
v-}(-u-\frac{3}{2})|_{c_1,c_2\rightarrow
\tilde{c}_1,\tilde{c}_2}, \label{ksk}
\end{equation}
where  $\tilde{c}_1$ and $\tilde{c}_2$ are the boundary parameters. For a generic choice of the four boundary parameters $\{c_i,\,\tilde{c}_i|i=1,2\}$,
it is easily to check that
$[K^{v-}(u),\,K^{v+}(v)]\neq 0$. This implies that the
$K^{\pm}(u)$ matrices  cannot be diagonalized simultaneously. In this case, it is quite hard to derive solutions via
conventional Bethe Ansatz methods
due to the absence of a proper reference state.
We will generalize the method developed in
Section 2 to obtain eigenvalues of the transfer matrix $t(u)$
(\ref{trweweu}) specified by the $K$-matrices (\ref{K-matrix-VV}) and
(\ref{ksk}) in the following subsections.

Following the method developed in \cite{Yan-06} and using the crossing-symmetry (\ref{Crossing-symmetry}) of the
$R$-matrix,  the explicit expressions (\ref{K-matrix-VV}) and (\ref{ksk}) of the
K-matrices, we find that the transfer matrix (\ref{trweweu}) possesses the crossing symmetry
\begin{eqnarray}
t(-u-\frac{3}{2})=t(u).\label{Transfer-Crossing}
\end{eqnarray}

\subsection{Operator product relations}

We define the dual fused monodromy matrices as
\begin{eqnarray}
&&\hat{T}_{\bar{0}}^{  \bar{v}}(u)=R^{    v\bar{v}}_{N\bar{0}}(u+\theta_N)\cdots R^{
v\bar{v}}_{2\bar{0}}(u+\theta_{2}) R^{
   v\bar{v}}_{1\bar{0}}(u+\theta_1), \no \\[4pt]
 &&\hat{T}_{\tilde {0}}^{  \tilde{v}}(u)=R^{    v\tilde{v}}_{N\tilde {0}}(u+\theta_N)\cdots R^{
v\tilde{v}}_{2\tilde {0}}(u+\theta_{2}) R^{
   v\tilde{v}}_{1\tilde {0}}(u+\theta_1),\no \\[4pt]
&&\hat{T}_{0}^{  s}(u)=R^{ vs}_{N{0}}(u+\theta_N)\cdots R^{
vs}_{2{0}}(u+\theta_{2}) R^{vs}_{1{0}}(u+\theta_1),\label{Mon-2}
\end{eqnarray}
where the $R$-matrices $R^{vs}_{21}(u)$ is defined by (\ref{Uni-SV-1}) and the others are defined by the  relations
\bea
&&R^{  \bar{v}v}_{\bar{1}2}(u)\,R^{v\bar{v}}_{2\bar{1}}(-u)=(u+1)(u-1)(u+\frac{3}{2})(u-\frac{3}{2})\times{\rm id}={\rho}_{\bar{v}}(u)\times{\rm id},\\
&&R^{
\tilde{v}v}_{\tilde{1}2}(u)\,R^{v\tilde{v}}_{2\tilde{1}}(-u)=(u+2)(u-2)(u+\frac{1}{2})(u-\frac{1}{2})\times{\rm
id}={\rho}_{\tilde{v}}(u)\times{\rm id}. \eea We have checked that the $R$-matrices also enjoy the properties
\begin{eqnarray}
&&R^{ \bar{v}v}_{\bar{1}2}(u)^{t_{\bar{1}}}R^{
v\bar{v}}_{2\bar{1}}(-u-3)^{t_{\bar{1}}}=u(u+3)(u+\frac{1}{2})(u+\frac{5}{2})\times{\rm
id}=\tilde{\rho}_{
\bar{v}}(u)\times{\rm id},\\[4pt]
&&R^{\tilde{v}v}_{\tilde{1}2}(u)^{t_{\tilde{1}}}R^{
v\bar{v}}_{2\tilde{1}}(-u-3)^{t_{\tilde{1}}}=(u+1)(u+2)(u-\frac{1}{2})(u+\frac{7}{2})\times{\rm
id}=\tilde{\rho}_{ \tilde{v}}(u)\times{\rm id},
\end{eqnarray}
which are used to derive the functional relations (\ref{futpp-7}) below.

Let us introduce  the  fused transfer matrices
\bea
&&t_2(u)=tr_{\bar{0}}\{K^{\bar{v}+}_{\bar{0}}(u) T_{\bar{0}}^{\bar{v}}(u)K^{\bar{v}-}_{\bar{0}}(u)\hat{T}_{\bar{0}}^{\bar{v}}(u)\},\no\\[4pt]
&&t_3(u)=tr_{\tilde{0}}\{K^{\tilde{v}+}_{\tilde{0}}(u) T_{\tilde{0}}^{\tilde{v}}(u)K^{\tilde{v}-}_{\tilde{0}}(u)\hat{T}_{\tilde{0}}^{\tilde{v}}(u)\},\no\\[4pt]
&&t_{s}(u)=tr_{0}\{K^{s+}_{0}(u) T_{0}^{  s}(u)K^{  s-}_{0}(u)\hat{T}_{0}^{  s}(u)\}, \label{2c0-1} \eea
where the associated $K$-matrices $K^{s\pm}(u)$, $K^{\bar{v}\pm}(u)$ and $K^{\tilde{v}\pm}(u)$ are given by (\ref{K-matwwwrix-VV})-(\ref{K-matwwwrix-wwVV}), (\ref{K-matrix-Bar-})-(\ref{K-matrix-Bar+}) and
(\ref{K-matrix-tilde-})-(\ref{K-matrix-tilde+}) respectively.
The Yang-Baxter relations, reflection equations and dual reflection equations allow one to show that all the transfer matrices constitute a commutative family \cite{Skl88}, namely,
\bea
[t_s(u),\,t_s(v)]=[t_s(u),\,t_m(v)]=[t_m(u),\,t_n(v)]=0,\quad m,n=1,2,3,
\eea
where we  have used the convention: $t_1(u)=t(u)$. Moreover, from the construction of the transfer matrices we know that $t_s(u)$, as a function of $u$, is a polynomial with degree $2N$, and that $t_1(u)$ is
a polynomial of $u$ with  degree $4N+2$, while $t_2(u)$ and $t_3(u)$ are polynomial of $u$ with degree $4N+4$. Thus one need to look for $14(N+1)$ independent conditions to completely determine
them.

Using the method we have used in previous sections and with the help of the relations (\ref{futt-7}) and (\ref{fuii-7}),  we can obtain
the operator product identities among the fused transfer matrices as\footnote{Due to the fact that for generic choices of inhomogeneous parameters $\{\theta_j\}$ the
identities (\ref{Op-pro-open-1})-(\ref{futpp-7})  involve the different points values of the transfer matrices, this implies that the identities should be  independent.}
\bea && t(\pm\theta_j)t(\pm\theta_j-\frac{3}{{2}})={ 2^{8}}
\frac{(\pm\theta_j-\frac{3}{{2}})(\pm\theta_j-\frac{1}{{4}})^3
(\pm\theta_j+\frac{3}{{2}})(\pm\theta_j+\frac{1}{{4}})^3}{(\pm\theta_j-\frac{3}{{4}})(\pm\theta_j-\frac{1}{{2}})
(\pm\theta_j+\frac{3}{{4}})(\pm\theta_j+\frac{1}{{2}})}\no\\[4pt]
&&\hspace{40mm}\times H(\pm\theta_j)H(\pm\theta_j-\frac{1}{{2}}), \label{Op-pro-open-1} \\[4pt]
&& t(\pm\theta_j)t(\pm\theta_j-1)= { 2^{2}} \frac{
(\pm\theta_j-1)(\pm\theta_j+\frac{3}{{2}})(\pm\theta_j+\frac{1}{{4}})^2}{(\pm\theta_j-\frac{1}{{2}})(\pm\theta_j-\frac{1}{{4}})
(\pm\theta_j+1)(\pm\theta_j+\frac{3}{{4}})}\nonumber\\[4pt]
&&\hspace{40mm}\times H(\pm\theta_j)t_2(\pm\theta_j-\frac{1}{{2}}), \no \\[4pt]
&& t(\pm\theta_j)t_2 (\pm\theta_j-\frac{3}{{2}})=-2^4
\frac{(\pm\theta_j-\frac{1}{{4}})(\pm\theta_j-\frac{3}{{2}})
(\pm\theta_j+\frac{3}{{2}})(\pm\theta_j+\frac{1}{{4}})^2}{(\pm\theta_j-\frac{1}{{2}})(\pm\theta_j+\frac{1}{{2}})
(\pm\theta_j+\frac{3}{{4}})}\nonumber\\[4pt]
&&\hspace{40mm}\times H(\pm\theta_j)
t_3(\pm\theta_j-1), \no \\[4pt]
&& t(\pm\theta_j)t_3(\pm\theta_j-2)= -{
2^{10}}{
(\pm\theta_j-1)(\pm\theta_j-2)(\pm\theta_j-\frac{3}{{2}})(\pm\theta_j+\frac{3}{{2}})(\pm\theta_j-\frac{1}{{4}})
}\nonumber\\[4pt]
&&\hspace{40mm}\times
\frac{(\pm\theta_j+\frac{1}{{4}})^2}{(\pm\theta_j+\frac{3}{{4}})
}H(\pm\theta_j)
t_{  s}(\pm\theta_j-\frac{1}{{4}})t_{  s}(\pm\theta_j-\frac{11}{{4}}), \no \\[4pt]
&& t(\pm\theta_j)t_2(\pm\theta_j-1)={ 2^{6}}
\frac{(\pm\theta_j-1)(\pm\theta_j-\frac{1}{{4}})(\pm\theta_j+\frac{1}{{4}})^4
(\pm\theta_j+\frac{3}{{2}})}{(\pm\theta_j-\frac{1}{{2}})(\pm\theta_j+1)
(\pm\theta_j+\frac{3}{{4}})}\nonumber\\[4pt]
&&\hspace{40mm}\times
 H(\pm\theta_j)t(\pm\theta_j-\frac{1}{{2}}),\no \\[4pt]
&& t(\pm\theta_j)t_3(\pm\theta_j-\frac{1}{{2}})= -{ 2^{4}} \frac{
(\pm\theta_j-\frac{1}{{4}})(\pm\theta_j+\frac{1}{{4}})^2}{(\pm\theta_j+\frac{3}{{4}})
} H(\pm\theta_j)t_2(\pm\theta_j),\no \\[4pt]
&& t(\pm\theta_j)t_{  s}(\pm\theta_j-\frac{5}{{4}})=-{
2^{4}} \frac{
(\pm\theta_j-\frac{1}{{4}})(\pm\theta_j+\frac{3}{{2}})(\pm\theta_j+\frac{1}{{4}})^2}{(\pm\theta_j-\frac{1}{{2}})(\pm\theta_j+\frac{3}{{4}})
} H(\pm\theta_j)t_{s}(\pm\theta_j-\frac{1}{{4}}),\label{futpp-7}
\eea
where \bea H(u)=(1+{c}_1{c}_2)(1+\tilde{c}_1\tilde{c}_2)\prod_{i=1}^N\tilde{\rho}_0(u-\theta_i)\tilde{\rho}_0(u+\theta_i). \no\eea
By direct calculation, we have
\bea &&t(0)= -\prod_{l=1}^N\rho_1
(-\theta_l)(1+c_1c_2)(1+\tilde{c}_1\tilde{c}_2)\,\times{\rm  id},\no \\[4pt]
&&t(-\frac{3}{2})=-\prod_{l=1}^N \rho_{
1}(-\theta_l)(1+c_1c_2)(1+\tilde{c}_1\tilde{c}_2)\times {\rm   id}, \no \\[4pt]
&&t_2(0)=-\frac{9}{4} \prod_{l=1}^N\rho_{ \bar v}
(-\theta_l)(1+c_1c_2)(1+\tilde{c}_1\tilde{c}_2)\,\times{\rm   id},\no \\[4pt]
&&t_2(-\frac{3}{2})=-\frac{9}{4}\prod_{l=1}^N \tilde{\rho}_{
\bar{v}}(-\theta_l-\frac{3}{2})(1+c_1c_2)(1+\tilde{c}_1\tilde{c}_2)\times {\rm
id}, \no \\[4pt]
 && t_2(-\frac{1}{2})= \frac{1}{4}\, t(-1),\quad t_2(-1)= \frac{1}{4} \,t(-\frac{1}{2}),\no \\[4pt]
&&t_3(0)= -\frac{45}{4}\prod_{l=1}^N\rho_{ \tilde v}
(-\theta_l)(1+c_1c_2)(1+\tilde{c}_1\tilde{c}_2)\times{\rm   id},\no \\[4pt]
&&t_3(-\frac{3}{2})=-\frac{45}{4}\prod_{l=1}^N \tilde{\rho}_{
\tilde{v}}(-\theta_l-\frac{3}{2})(1+c_1c_2)(1+\tilde{c}_1\tilde{c}_2)\times {\rm  id}, \no \\[4pt]
 && t_3(-1)= \frac{1}{3}\, t_2(-\frac{3}{2}),\quad  t_3(-\frac{1}{2})= \frac{3}{4}\,t(-\frac{3}{2}),\label{t3-4}
 \eea
where we have used the identities:
\bea &&tr \{K^{  v
+}(0)\}=1+\tilde{c}_1\tilde{c}_2,\hspace{2mm}K^{
v-}(0)=-(1+c_1c_2)\times {\rm id }, \no\\[4pt]
&&tr\{K^{
v-}(-\frac{3}{2})\}=1+c_1c_2,\hspace{2mm}K^{v+}(-\frac{3}{2})=-(1+\tilde{c}_1\tilde{c}_2)\times {\rm id },\no\\[4pt]
&&tr \{K^{  \bar{v}
+}(0)\}=-\frac{3}{2}(1+\tilde{c}_1\tilde{c}_2),\hspace{2mm}K^{
\bar{v}-}(0)=\frac{3}{2}(1+c_1c_2)\times {\rm id },\no\\[4pt]
&&tr\{K^{
\bar{v}-}(-\frac{3}{2})\}=-\frac{3}{2}(1+c_1c_2),\hspace{2mm}K^{
\bar{v}+}(-\frac{3}{2})=\frac{3}{2}(1+\tilde{c}_1\tilde{c}_2)\times {\rm id },\no\\[4pt]
&&tr \{K^{  \tilde{v}
+}(0)\}=\frac{15}{2}(1+\tilde{c}_1\tilde{c}_2),\hspace{2mm}K^{
\tilde{v}-}(0)=-\frac{3}{2}(1+c_1c_2)\times {\rm id },\no\\[4pt]
&&tr\{K^{
\tilde{v}-}(-\frac{3}{2})\}=\frac{15}{2}(1+c_1c_2),\hspace{2mm}K^{
\tilde{v}+}(-\frac{3}{2})=-\frac{3}{2}(1+\tilde{c}_1\tilde{c}_2)\times {\rm id },\no\\[4pt]
&&tr_1\{R^{  vv}_{12}(-1)K^{  v+}_1(0)R^{
vv}_{21}(-2)\}=\frac{3}{2}(1+\tilde{c}_1\tilde{c}_2)\times {\rm id },\no\\[4pt]
&&tr_2\{R^{  vv}_{21}(-2)K^{  v-}_2(-\frac{3}{2})R^{
 vv}_{12}(-1)\}^{t_1t_2}=\frac{3}{2}(1+c_1c_2)\times {\rm id },\no\\[4pt]
 &&tr_1\{R^{  vv}_{12}(-1)R^{  vv}_{13}(-2)K^{  v+}_1(0)R^{  vv}_{31}(-1)R^{  vv}_{21}(-2)\}=
 \frac{9}{4}(1+\tilde{c}_1\tilde{c}_2)\times {\rm id },\no\\[4pt]
 && tr_{23}\{P_{123}R^{  vv}_{23}(-1)R^{
vv}_{13}(-2)K^{  v+}_2(-\frac{1}{2})R^{  vv}_{12}(-3)K^{
v+}_1(\frac{1}{2}) R^{  vv}_{32}(-1)R^{ vv}_{31}(-2)R^{
vv}_{21}(0)\}\no\\[4pt]
&&\quad\quad\quad\quad=-\frac{3^5}{4}(1+\tilde{c}_1\tilde{c}_2)^2\times {\rm id },\no\\[4pt]
 &&K^{  v-}(-\frac{1}{2})K^{  v-}(\frac{1}{2})=-3(1+c_1c_2)^2\times {\rm id }.\no\eea
Moreover, we can derive the asymptotic behaviors of
the transfer matrices
\bea && t(u)|_{u\rightarrow \pm\infty}=  -4[(4+2c_1\tilde{c}_2+2c_2\tilde{c}_1)^2+4(1+{c}_1{c}_2)(1+\tilde{c}_1\tilde{c}_2)]  u^{4N+2}\times {\rm id}+\cdots, \no \\[4pt]
&& t_2(u)|_{u\rightarrow \pm\infty}=  2^{6}[\frac{3}{4}(4+2c_1\tilde{c}_2+2c_2\tilde{c}_1)^2-(1+{c}_1{c}_2)(1+\tilde{c}_1\tilde{c}_2)] u^{4N+4}\times{\rm id}+\cdots,\no \\[4pt]
&&t_3(u)|_{u\rightarrow \pm\infty}=
2^{6}[(4+2c_1\tilde{c}_2+2c_2\tilde{c}_1)^2-(1+{c}_1{c}_2)(1+\tilde{c}_1\tilde{c}_2)]
u^{4N+4}\times {\rm id}+\cdots, \no \\[4pt]
&& t_{s}(u)|_{u\rightarrow \pm\infty}= (4+2c_1\tilde{c}_2+2c_2\tilde{c}_1) u^{2N} \times{\rm id} +\cdots. \label{as-tv2} \eea

Let us denote the eigenvalues of the fused transfer matrices $t(u)$, $t_2(u)$ $t_3(u)$ and $t_s(u)$ as
$\Lambda(u)$, $\Lambda_2(u)$, $\Lambda_3(u)$ and $\Lambda_s(u)$, respectively.
The above $14(N+1)$ operator production identities (\ref{Op-pro-open-1})-(\ref{as-tv2}) directly imply that
\bea && \Lambda(\pm\theta_j)\Lambda(\pm\theta_j-\frac{3}{{2}})={ 2^{8}}
\frac{(\pm\theta_j-\frac{3}{{2}})(\pm\theta_j-\frac{1}{{4}})^3
(\pm\theta_j+\frac{3}{{2}})(\pm\theta_j+\frac{1}{{4}})^3}{(\pm\theta_j-\frac{3}{{4}})(\pm\theta_j-\frac{1}{{2}})
(\pm\theta_j+\frac{3}{{4}})(\pm\theta_j+\frac{1}{{2}})}\nonumber\\[4pt]
&&\hspace{40mm}\times H(\pm\theta_j)H(\pm\theta_j-\frac{1}{{2}}), \label{Eigen-open-1} \\[4pt]
&& \Lambda(\pm\theta_j)\Lambda(\pm\theta_j-1)= {
2^{2}} \frac{
(\pm\theta_j-1)(\pm\theta_j+\frac{3}{{2}})(\pm\theta_j+\frac{1}{{4}})^2}{(\pm\theta_j-\frac{1}{{2}})(\pm\theta_j-\frac{1}{{4}})
(\pm\theta_j+1)(\pm\theta_j+\frac{3}{{4}})}\nonumber\\[4pt]
&&\hspace{40mm}\times H(\pm\theta_j)\Lambda_2(\pm\theta_j-\frac{1}{{2}}),  \\[4pt]
&& \Lambda(\pm\theta_j)\Lambda_2(\pm\theta_j-\frac{3}{{2}})=-2^4
\frac{(\pm\theta_j-\frac{1}{{4}})(\pm\theta_j-\frac{3}{{2}})
(\pm\theta_j+\frac{3}{{2}})(\pm\theta_j+\frac{1}{{4}})^2}{(\pm\theta_j-\frac{1}{{2}})(\pm\theta_j+\frac{1}{{2}})
(\pm\theta_j+\frac{3}{{4}})}\nonumber\\[4pt]
&&\hspace{40mm}\times H(\pm\theta_j)
\Lambda_3(\pm\theta_j-1),  \\[4pt]
&& \Lambda(\pm\theta_j)\Lambda_3 (\pm\theta_j-2)= -{ 2^{10}}{
(\pm\theta_j-1)(\pm\theta_j-2)(\pm\theta_j-\frac{3}{{2}})(\pm\theta_j+\frac{3}{{2}})(\pm\theta_j-\frac{1}{{4}})
}\nonumber\\[4pt]
&&\hspace{40mm}\times
\frac{(\pm\theta_j+\frac{1}{{4}})^2}{(\pm\theta_j+\frac{3}{{4}})
}H(\pm\theta_j)
\Lambda_{ s}(\pm\theta_j-\frac{1}{{4}})\Lambda_{  s}(\pm\theta_j-\frac{11}{{4}}),  \\[4pt]
&& \Lambda(\pm\theta_j)\Lambda_2(\pm\theta_j-1)={ 2^{6}}
\frac{(\pm\theta_j-1)(\pm\theta_j-\frac{1}{{4}})(\pm\theta_j+\frac{1}{{4}})^4
(\pm\theta_j+\frac{3}{{2}})}{(\pm\theta_j-\frac{1}{{2}})(\pm\theta_j+1)
(\pm\theta_j+\frac{3}{{4}})}\nonumber\\[4pt]
&&\hspace{40mm}\times
 H(\pm\theta_j)\Lambda(\pm\theta_j-\frac{1}{{2}}), \\[4pt]
&& \Lambda(\pm\theta_j)\Lambda_3 (\pm\theta_j-\frac{1}{{2}})= -{ 2^{4}} \frac{
(\pm\theta_j-\frac{1}{{4}})(\pm\theta_j+\frac{1}{{4}})^2}{(\pm\theta_j+\frac{3}{{4}})
} H(\pm\theta_j)\Lambda_2(\pm\theta_j), \\[4pt]
&& \Lambda(\pm\theta_j)\Lambda_{
s}(\pm\theta_j-\frac{5}{{4}})=-{ 2^{4}} \frac{
(\pm\theta_j-\frac{1}{{4}})(\pm\theta_j+\frac{3}{{2}})(\pm\theta_j+\frac{1}{{4}})^2}{(\pm\theta_j-\frac{1}{{2}})(\pm\theta_j+\frac{3}{{4}})
}\nonumber\\[4pt]
&&\hspace{40mm}\times
H(\pm\theta_j)\Lambda_{s}(\pm\theta_j-\frac{1}{{4}}),\label{feutpp-7}\\[4pt]
&&\Lambda(0)= -\prod_{l=1}^N\rho_1
(-\theta_l)(1+c_1c_2)(1+\tilde{c}_1\tilde{c}_2), \\[4pt]
&&\Lambda(-\frac{3}{2})=-\prod_{l=1}^N \rho_{
1}(-\theta_l)(1+c_1c_2)(1+\tilde{c}_1\tilde{c}_2),  \\[4pt]
&&\Lambda_2(0)= -\frac{9}{4}\prod_{l=1}^N\rho_{ \bar v}
(-\theta_l)(1+c_1c_2)(1+\tilde{c}_1\tilde{c}_2), \\[4pt]
&&\Lambda_2(-\frac{3}{2})=-\frac94\prod_{l=1}^N \tilde{\rho}_{
\bar{v}}(-\theta_l-\frac{3}{2})(1+c_1c_2)(1+\tilde{c}_1\tilde{c}_2), \\[4pt]
&& \Lambda_2(-\frac{1}{2})= \frac14\Lambda(-1),  \quad\Lambda_2(-1)= \frac14\Lambda(-\frac{1}{2}), \\[4pt]
&&\Lambda_3(0)= -\frac{45}{4}\prod_{l=1}^N\rho_{ \tilde v}
(-\theta_l)(1+c_1c_2)(1+\tilde{c}_1\tilde{c}_2), \\[4pt]
&&\Lambda_3(-\frac{3}{2})=-\frac{45}{4}\prod_{l=1}^N
\tilde{\rho}_{ \tilde{v}}(-\theta_l-\frac{3}{2})(1+c_1c_2)(1+\tilde{c}_1\tilde{c}_2), \\[4pt]
 && \Lambda_3(-1)= \frac13\Lambda_2(-\frac{3}{2}), \quad \Lambda_3(-\frac{1}{2})= \frac34\Lambda(-\frac{3}{2}),\\[4pt]
&& \Lambda(u)|_{u\rightarrow \pm\infty}=  -4[(4+2c_1\tilde{c}_2+2c_2\tilde{c}_1)^2+4(1+{c}_1{c}_2)(1+\tilde{c}_1\tilde{c}_2)]  u^{4N+2}+\cdots,  \\[4pt]
&& \Lambda_2(u)|_{u\rightarrow \pm\infty}=  2^{6}[\frac{3}{4}(4+2c_1\tilde{c}_2+2c_2\tilde{c}_1)^2-(1+{c}_1{c}_2)(1+\tilde{c}_1\tilde{c}_2)] u^{4N+4}+\cdots, \\[4pt]
&&\Lambda_3(u)|_{u\rightarrow \pm\infty}=
2^{6}[(4+2c_1\tilde{c}_2+2c_2\tilde{c}_1)^2-(1+{c}_1{c}_2)(1+\tilde{c}_1\tilde{c}_2)]
u^{4N+4}+\cdots,  \\[4pt]
&& \Lambda_{s}(u)|_{u\rightarrow \pm\infty}=
(4+2c_1\tilde{c}_2+2c_2\tilde{c}_1) u^{2N}  +\cdots.
\label{Eigen-open-2} \eea
The above $14(N+1)$ conditions (\ref{Eigen-open-1})-(\ref{Eigen-open-2}) could completely
determine the eigenvalues $\Lambda(u)$, $\Lambda_2(u)$,
$\Lambda_3(u)$ and $\Lambda_s(u)$
in terms of some inhomogeneous $T-Q$ relations.

\subsection{Inhomogeneous $T-Q$ relations}

Define
 \bea &&Z_1(u)=-{ 2^{4}} \frac{
(u+\frac{1}{{4}})^3(u+\frac{3}{{2}})}{(u+\frac{1}{{2}})(u+\frac{3}{{4}})
}\prod_{j=1}^N
a_1(u-\theta_j)a_1(u+\theta_j)\frac{Q^{(1)}(u-1)}{Q^{(1)}(u)}\no\\[4pt]
&&\hspace{20mm}\times (1+{c}_1{c}_2)(1+\tilde{c}_1\tilde{c}_2),\no\\[4pt]
&&Z_2(u)=-{ 2^{4}} \frac{
u(u+\frac{3}{{2}})(u+\frac{1}{{4}})(u+\frac{3}{{4}})}{(u+1)(u+\frac{1}{{2}})
}\prod_{j=1}^N
b_1(u-\theta_j)b_1(u+\theta_j)\frac{Q^{(1)}(u+1)Q^{(2)}(u-1)}{Q^{(1)}(u)Q^{(2)}(u)}\no\\[4pt]
&&\hspace{20mm}\times (1+{c}_1{c}_2)(1+\tilde{c}_1\tilde{c}_2),\no\\[4pt]
&&Z_3(u)=-{ 2^{4}}\frac{
u(u+\frac{3}{{2}})(u+\frac{1}{{4}})(u+\frac{5}{{4}})}{(u+1)(u+\frac{1}{{2}})
}\prod_{j=1}^N
b_1(u-\theta_j)b_1(u+\theta_j)\frac{Q^{(2)}(u-1)Q^{(2)}(u+\frac{1}{2})}{Q^{(2)}(u)Q^{(2)}(u-\frac{1}{2})}\no\\[4pt]
&&\hspace{20mm}\times (1+{c}_1{c}_2)(1+\tilde{c}_1\tilde{c}_2),\no\\[4pt]
&&Z_4(u)=-{ 2^{4}}\frac{
u(u+\frac{3}{{2}})(u+\frac{3}{{4}})(u+\frac{5}{{4}})}{(u+1)(u+\frac{1}{{2}})
}\prod_{j=1}^N
b_1(u-\theta_j)b_1(u+\theta_j)\frac{Q^{(1)}(u-\frac{1}{2})Q^{(2)}(u+\frac{1}{2})}{Q^{(1)}(u+\frac{1}{2})Q^{(2)}(u-\frac{1}{2})}\no\\[4pt]
&&\hspace{20mm}\times (1+{c}_1{c}_2)(1+\tilde{c}_1\tilde{c}_2),\no\\[4pt]
&&Z_5(u)=-{ 2^{4}} \frac{
u(u+\frac{5}{{4}})^3}{(u+1)(u+\frac{3}{{4}}) }\prod_{j=1}^N
e_1(u-\theta_j)e_1(u+\theta_j)
\frac{Q^{(1)}(u+\frac{3}{2})}{Q^{(1)}(u+\frac{1}{2})}\no\\[4pt]
&&\hspace{20mm}\times
(1+{c}_1{c}_2)(1+\tilde{c}_1\tilde{c}_2),\no\\[4pt]
&&Q^{(m)}(u)=\prod_{k=1}^{L_m}(u-\lambda_k^{(m)}+\frac{m}{2})(u+\lambda_k^{(m)}+\frac{m}{2}), \quad m=1,2, \eea
and
 \bea &&f_1(u)=-{ 2^{4}} \frac{u
(u+\frac{3}{{2}})(u+\frac{1}{{4}})(u+\frac{3}{{4}})(u+\frac{5}{{4}})}{(u+\frac{1}{{2}})
}\prod_{j=1}^N
[b_1(u-\theta_j)b_1(u+\theta_j)(u-\theta_j+1)\no\\[4pt]
&&\hspace{10mm}\times (u+\theta_j+1)]\frac{Q^{(1)}(u+1)Q^{(1)}(u+\frac{1}{{2}})Q^{(2)}(u-1)}{Q^{(2)}(u)Q^{(2)}(u-\frac{1}{{2}})}
(1+{c}_1{c}_2)(1+\tilde{c}_1\tilde{c}_2)\,x,\no\\[4pt]
&&f_2(u)=-{ 2^{4}} \frac{
u(u+\frac{3}{{2}})(u+\frac{3}{{4}})^2(u+\frac{5}{{4}})}{(u+\frac{1}{{2}})
}\prod_{j=1}^N
[b_1(u-\theta_j)b_1(u+\theta_j)(u-\theta_j+1)\no\\[4pt]
&&\hspace{10mm}\times (u+\theta_j+1)] \frac{Q^{(1)}(u-\frac{1}{{2}})Q^{(1)}(u+1)}{Q^{(2)}(u-\frac{1}{{2}})}
(1+{c}_1{c}_2)(1+\tilde{c}_1\tilde{c}_2)x,\no\\[4pt]
&&f_3(u)=-{ 2^{4}} {
u(u+\frac{3}{{2}})}{(u+\frac{1}{{4}})(u+\frac{3}{{4}})^2(u+\frac{5}{{4}})
}\prod_{j=1}^N
a_1(u-\theta_j)a_1(u+\theta_j)e_1(u-\theta_j)
\no\\[4pt]
&&\hspace{10mm}\times e_1(u+\theta_j) \frac{Q^{(1)}(u)Q^{(1)}(u+\frac{1}{{2}})Q^{(1)}(u+1)Q^{(1)}(u-\frac{1}{{2}})}{Q^{(2)}(u)Q^{(2)}(u-\frac{1}{2})}
(1+{c}_1{c}_2)(1+\tilde{c}_1\tilde{c}_2)x^2,\no\\[4pt]
&&f_4(u)=-{ 2^{4}} \frac{
u(u+\frac{3}{{2}})(u+\frac{1}{{4}})(u+\frac{3}{{4}})^2}{(u+1)
}\prod_{j=1}^N
[b_1(u-\theta_j)b_1(u+\theta_j)(u-\theta_j+\frac{1}{2})\no\\[4pt]
&&\hspace{10mm}\times (u+\theta_j+\frac{1}{2})] \frac{Q^{(1)}(u+1)Q^{(1)}(u-\frac{1}{{2}})}{Q^{(2)}(u)}(1+{c}_1{c}_2)(1+\tilde{c}_1\tilde{c}_2)\,x,\no\\[4pt]
&&f_5(u)=-{ 2^{4}} \frac{u
(u+\frac{3}{{2}})(u+\frac{1}{{4}})(u+\frac{3}{{4}})(u+\frac{5}{{4}})}{(u+1)
}\prod_{j=1}^N
[b_1(u-\theta_j)b_1(u+\theta_j)(u-\theta_j+\frac{1}{2})\no\\[4pt]
&&\hspace{10mm}\times(u+\theta_j+\frac{1}{2})]
\frac{Q^{(2)}(u+\frac{1}{{2}})Q^{(1)}(u)Q^{(1)}(u-\frac{1}{{2}})}{Q^{(2)}(u)Q^{(2)}(u-\frac{1}{{2}})}
 (1+{c}_1{c}_2)(1+\tilde{c}_1\tilde{c}_2)\,x, \label{function}
\eea
where $x$ is a parameter which will be determined later (see (\ref{x-value}) below). The $14(N+1)$ functional relations (\ref{Eigen-open-1})-(\ref{Eigen-open-2})
allow us to express the eigenvalues  of the transfer matrices in terms of some inhomogeneous $T-Q$ relations as follows:
\bea
&&\Lambda(u)=Z_1(u)+
Z_2(u)+Z_3(u)+Z_4(u)+Z_5(u)\no\\[4pt]
&&\qquad\qquad +f_1(u)+f_2(u)+f_3(u)+f_4(u)+f_5(u),\label{Eigen-open-Lambda} \\[4pt]
&&\Lambda_2(u) =2^{-6}[(u-\frac12)(u+2)(u+\frac{3}{4})^2H(u+\frac12)]^{-1}\tilde{\rho}_{v}(2u) \no \\[4pt]
&&\qquad \times \{
Z_1(u+\frac12)[Z_2(u-\frac12)+Z_3(u-\frac12)+Z_4(u-\frac12)+f_1(u-\frac12)+f_2(u-\frac12)
\no\\[4pt]
&& \qquad
+f_3(u-\frac12)+f_4(u-\frac12)+f_5(u-\frac12)]+[Z_2(u+\frac12)+Z_3(u+\frac12)+Z_4(u+\frac12)\no\\[4pt]
&& \qquad +f_1(u+\frac12)+f_2(u+\frac12)+f_3(u+\frac12)+f_4(u+\frac12)+f_5(u+\frac12)]Z_5(u-\frac12)
\no\\[4pt]
&& \qquad
+[Z_2(u+\frac12)+Z_3(u+\frac12)+f_1(u+\frac12)][Z_3(u-\frac12)\no\\[4pt]
&&\qquad  +Z_4(u-\frac12)+f_5(u-\frac12)]+Z_1(u+\frac12)Z_5(u-\frac12)\}, \no  \\[4pt]
&&
\Lambda_3(u) =-2^{-18}[(u+\frac{5}{{4}})^3(u+\frac{3}{{4}})^2(u-\frac{1}{{2}})u(u-1)(u+\frac{1}{{4}})^3(u+\frac{3}{{2}})\nonumber\\[4pt]
&&\qquad \times
(u+\frac{5}{{2}})(u+2) H(u+1)H(u)]^{-1}\tilde{\rho}_{ v} (2u+1)\tilde{\rho}_{  v}
(2u)\tilde{\rho}_{  v} (2u-1)
 \no\\[4pt]
&&\qquad \times\{[Z_1(u+1)Z_2(u)+Z_1(u+1)Z_3(u)+Z_2(u+1)Z_3(u)\no\\[4pt]
&&\qquad+Z_3(u+1)Z_3(u)+Z_1(u+1)f_1(u)+f_1(u+1)Z_3(u)][Z_3(u-1)+Z_4(u-1)\no\\[4pt]
&&\qquad+f_5(u-1)]+Z_1(u+1)[Z_2(u)+Z_3(u)+Z_4(u)+f_1(u)+f_2(u) \no\\[4pt]
&&\qquad+f_3(u)+f_4(u)+f_5(u)]Z_5(u-1)+[Z_2(u+1)+Z_3(u+1)\no\\[4pt]
&&\qquad+f_1(u+1)][Z_3(u)+Z_4(u)+f_5(u)]Z_5(u-1)\}, \no \\[4pt]
&&\Lambda_{s}(u)=\frac{1}{2^8}\frac{
(u-\frac{1}{{4}})(u+1)[(1+{c}_1{c}_2)(1+\tilde{c}_1\tilde{c}_2)]^{-\frac{3}{{2}}}}{(u+\frac{3}{{4}})(u+\frac{7}{{4}})(u-\frac{3}{{4}})(u-\frac{1}{{2}})(u+\frac{1}{{2}})^2
}\frac{Q^{(2)}(u-\frac{5}{4})}{Q^{(2)}(u-\frac{7}{4})}\no\\[4pt]
&& \qquad \times \prod_{j=1}^N
[(u+\frac{7}{{4}}-\theta_j)(u+\frac{7}{{4}}+\theta_j)b_1(u-\frac{3}{{4}}-\theta_j)b_1(u-\frac{3}{{4}}+\theta_j)]^{-1}
\no\\[4pt]
&&\qquad \times\{Z_1(u+\frac{1}{{4}})Z_2(u-\frac{3}{{4}})+Z_1(u+\frac{1}{{4}})Z_3(u-\frac{3}{{4}})+Z_2(u+\frac{1}{{4}})Z_3(u-\frac{3}{{4}})\no\\[4pt]
&&\qquad  +Z_3(u+\frac{1}{{4}})Z_3(u-\frac{3}{{4}})+Z_1(u+\frac{1}{{4}})f_1(u-\frac{3}{{4}})+f_1(u+\frac{1}{{4}})Z_3(u-\frac{3}{{4}})\} \no \\[4pt]
&&\quad =\frac{1}{2^8}\frac{
(u+\frac{1}{{2}})(u+\frac{7}{{4}})[(1+{c}_1{c}_2)(1+\tilde{c}_1\tilde{c}_2)]^{-\frac{3}{{2}}}}{(u-\frac{1}{{4}})(u+\frac{3}{{4}})(u+2)(u+\frac{9}{{4}})(u+1)^2
}\frac{Q^{(2)}(u+\frac{3}{4})}{Q^{(2)}(u+\frac{5}{4})}\no\\[4pt]
&& \qquad \times \prod_{j=1}^N
[(u-\frac{1}{{4}}-\theta_j)(u-\frac{1}{{4}}+\theta_j)b_1(u+\frac{3}{{4}}-\theta_j)b_1(u+\frac{3}{{4}}+\theta_j)]^{-1}
\no\\[4pt]
&&\qquad \times\{Z_3(u+\frac{3}{{4}})Z_3(u-\frac{1}{{4}})+Z_3(u+\frac{3}{{4}})Z_4(u-\frac{1}{{4}})+Z_3(u+\frac{3}{{4}})Z_5(u-\frac{1}{{4}})\no\\[4pt]
&&\qquad+Z_4(u+\frac{3}{{4}})Z_5(u-\frac{1}{{4}})+Z_3(u+\frac{3}{{4}})f_5(u-\frac{1}{{4}})+f_5(u+\frac{3}{{4}})Z_5(u-\frac{1}{{4}})\}. \label{eop-2321}
\eea
All the eigenvalues are polynomials of $u$, the residues of
right hand sides of Eq.(\ref{Eigen-open-Lambda}) should be
zero, which gives rise to the BAEs
\bea
&&\frac{Q^{(1)}(\lambda_k^{(1)}+\frac{1}{2})Q^{(2)}(\lambda_k^{(1)}-\frac{3}{2})}{Q^{(1)}(\lambda_k^{(1)}-\frac{3}{2})Q^{(2)}(\lambda_k^{(1)}-\frac{1}{2})}
=-\frac{(\lambda_k^{(1)}+\frac{1}{2})(\lambda_k^{(1)}-\frac{1}{4})^2}{(\lambda_k^{(1)}-\frac{1}{2})(\lambda_k^{(1)}+\frac{1}{4})^2}\no\\[4pt]
&& \qquad\qquad \times
\frac{\prod_{j=1}^N(\lambda_k^{(1)}-\theta_j+\frac{1}{2})
(\lambda_k^{(1)}+\theta_j+\frac{1}{2})}{\prod_{j=1}^N(\lambda_k^{(1)}-\theta_j-\frac{1}{2})(\lambda_k^{(1)}+\theta_j-\frac{1}{2})},\quad k=1,2,\cdots,L_1, \label{opba-1} \\[8pt]
&&\frac{(\lambda_l^{(2)}-\frac{1}{4})}{(\lambda_l^{(2)}+\frac{1}{4})}
\frac{Q^{(2)}(\lambda_l^{(2)}-\frac{3}{2})}{Q^{(1)}(\lambda_l^{(2)}-1)Q^{(1)}(\lambda_l^{(2)}-\frac{1}{2})}
+
\frac{Q^{(2)}(\lambda_l^{(2)}-\frac{1}{2})}{Q^{(1)}(\lambda_l^{(2)})Q^{(1)}(\lambda_l^{(2)}-\frac{1}{2})}\no\\[4pt]
&&\qquad\qquad=-x\lambda_l^{(2)}(\lambda_l^{(2)}-\frac{1}{4})\prod_{j=1}^N(\lambda_l^{(2)}-\theta_j)(\lambda_l^{(2)}+\theta_j),
\quad l=1,2,\cdots,L_2. \label{opba-2}\eea
Considering the asymptotic behaviors of $\Lambda(u)$, $\Lambda_2(u)$, $\Lambda_3(u)$ and $\Lambda_s(u)$,
we obtain a constraint between the two no-negative integers $L_1$ and $L_2$:
\bea
L_2=2L_1+N+1,\label{L1-L2-Constraint}
\eea
which implies that only one of the two $U(1)$ conserved charges (related to the Cartan sub-algebra of $so(5)$) has been broken by the boundary
terms in (\ref{hh}). Moreover,
the value of parameter $x$ in the functions $\{f_m(u)\}$ given in (\ref{function}) as
\bea
&&x=\frac{4+2c_1\tilde{c}_2+2c_2\tilde{c}_1}{2\sqrt{(1+{c}_1{c}_2)(1+\tilde{c}_1\tilde{c}_2)}}-2.\label{x-value} \eea

Some remarks are in order. The expression of the inhomogeneous $T-Q$ relation (\ref{Eigen-open-Lambda}) and the associated BAEs (\ref{opba-1})-
(\ref{opba-2}) imply  that one of the two $U(1)$ conserved charges (corresponding to the non-negative integer $L_1$) is still survived,
which means that $L_1$ can take some integers. Following the method developed in \cite{Sun-17} and combining the results in \cite{b2aba}, we know that
$0\leq L_1\leq 2N$ and then $L_2$ is fixed by the constraint (\ref{L1-L2-Constraint}), which would give complete spectrum of the transfer matrix (\ref{trweweu}).
However, for more generic
$K$-matrices the two $U(1)$ charges might be all broken, then both non-negative integers $L_1$ and $L_2$ will be fixed\footnote{It has been confirmed
for the $B_1$ model with generic boundary terms \cite{Cao15JHEP036, Xue19}.}.

We have checked that the  BAEs (\ref{opba-1})-(\ref{opba-2}) indeed ensure that all the $T-Q$ relations (\ref{Eigen-open-Lambda})-(\ref{eop-2321}) are  polynomials of $u$ with the required degrees and also  satisfy the required $14(N+1)$ function relations (\ref{Eigen-open-1})-(\ref{Eigen-open-2}).
Therefore, we conclude that $\Lambda(u)$, $\Lambda_2(u)$, $\Lambda_3(u)$ and $\Lambda_s(u)$ given by the inhomogeneous $T-Q$ relations (\ref{Eigen-open-Lambda})-(\ref{eop-2321}) are
indeed the eigenvalues of  the transfer matrices $t(u)$, $t_2(u)$, $t_3(u)$ and $t_s(u)$ provided that the $L_1+L_2$ parameters $\{\lambda^{(m)}_k\}$ satisfy the associated BAEs  (\ref{opba-1})-(\ref{opba-2}). Moreover, when $c_1=c_2=\tilde{c}_1=\tilde{c}_2=0$, the boundary reflection matrices degenerate to diagonal ones and our results recover those
obtained by the algebraic Bethe method \cite{b2aba}.

\section{Discussion}

In this paper, we study the $so(5)$ quantum spin chains with integrable boundary conditions.
By using the fusion technique, we obtain the closed operator product identities of the fused transfer matrices (\ref{Op-Product-Periodic-1})-(\ref{futp-7})
for the periodic boundary condition and (\ref{Op-pro-open-1})-(\ref{futpp-7}) for the off-diagonal boundary case. Based on them and the asymptotic
behaviors as well as the special values at certain points, we obtain the exact solutions of the systems. For the periodic case,
the eigenvalues of the transfer matrices are given in terms of the homogeneous $T-Q$ relations (\ref{T-Q-Hom-1})-(\ref{ep-2}) and
 the associated BAEs are (\ref{BAEs-1})-(\ref{BAEs-2}), which recover those obtained previously via conventional Bethe Ansatz methods\cite{NYReshetikhin1,NYReshetikhin2,Bn}. For the open boundary case specified by the off-diagonal boundary $K$-matrices (\ref{K-matrix-VV})-(\ref{ksk}), (\ref{K-matwwwrix-VV})-(\ref{K-matwwwrix-wwVV}), (\ref{K-matrix-Bar-})-(\ref{K-matrix-Bar+}) and
(\ref{K-matrix-tilde-})-(\ref{K-matrix-tilde+}), the eigenvalues of the transfer matrices are given in terms of the inhomogeneous $T-Q$ relations (\ref{Eigen-open-Lambda})-(\ref{eop-2321}) and the associated BAEs (\ref{opba-1})-(\ref{opba-2}).  The method and the results in this paper can be generalized to the  $so(2n+1)$ (i.e., $B_n$) case directly.

\section*{Acknowledgments}

The financial supports from the National Program for Basic
Research of MOST (Grant Nos. 2016YFA0300600 and 2016YFA0302104),
the National Natural Science Foundation of China (Grant Nos.
11434013, 11425522, 11547045, 11774397, 11775178, 11775177 and
91536115), the Major Basic Research Program of Natural Science of
Shaanxi Province (Grant Nos. 2017KCT-12, 2017ZDJC-32), Australian Research Council (Grant No. DP 190101529)
and the Strategic Priority Research Program of the Chinese Academy of
Sciences, and the Double First-Class University Construction
Project of Northwest University are gratefully acknowledged. GL Li
acknowledges support from Shaanxi Province Key Laboratory of Quantum
Information and Quantum Optoelectronic Devices, Xian Jiaotong
University. We would like to thank Prof. R. Nepomechie for drawing our
attention on the crossing-symmetry (\ref{Crossing-symmetry}) of the $R$-matrix which
enables us to prove the crossing symmetry (\ref{Transfer-Crossing}) of the fundamental transfer matrix $t(u)$.

\section*{Appendix A: Projectors and the fusion of the $R$-matrices}
\setcounter{equation}{0}
\renewcommand{\theequation}{A.\arabic{equation}}

\subsection*{A.1: Fused $R$-matrices}
From the definition (\ref{Fused-R-1}) of the fused $R$-matrix $R_{\bar{1}2}^{  \bar{v}v}(u)$, we have
\bea R_{\bar{1}2}^{  \bar{v}v}(-1)=
P^{{  \bar{v}v}(5) }_{\bar{1}2}\times \bar{S},\no \eea
where $P^{{  \bar{v}v}(5) }_{\bar{1}2}$  is a $5$-dimensional projector (see  (\ref{a6}) below). The new projector allows us to construct
a new fused $R$-matrix,
\bea
&&R_{\langle\bar{1}2\rangle3}^{  vv}(u)=\tilde{\rho}_0^{-1}(u+\frac{1}{2}) P^{{
\bar{v}v}(5) }_{\bar{1}2}R^{  vv}_{23}(u+\frac{1}{2})R^{
\bar{v}v}_{\bar{1}3}(u-\frac{1}{2})P^{{  \bar{v}v}(5)}_{\bar{1}2}. \eea
Taking the correspondence
\bea
|\phi^{(5)}_i\rangle\longrightarrow |i\rangle,\quad i=1,\cdots,5,\label{Identification-1}
\eea
where the basis $\{|\phi^{(5)}_i\rangle\}$ is given by (\ref{Basis-5}) below, we have
\bea
R_{\langle\bar{1}2\rangle3}^{  vv}(u)\equiv R^{vv}_{13}(u).\label{R5-5}
\eea
The definition (\ref{Fused-R-1}) of the fused $R$-matrix $R_{\tilde{1}2}^{  \tilde{v}v}(u)$ implies that
\bea &&R_{\tilde{1}2}^{  \tilde{v}v}(-\frac{1}{2})=
P^{{  \tilde{v}v}(11) }_{\tilde{1}2}\times \tilde{S}, \eea
where $P^{{  \tilde{v}v}(11) }_{\tilde{1}2}$ is a $11$-dimensional projector (see below (\ref{a8})) and
$\tilde{S}$ is an irrelevant constant matrix. The projector allows us to construct the fused $R$-matrix
\bea &&R_{\langle \tilde{1}2\rangle 3}^{
\bar{v}v}(u)=\tilde{\rho}_0^{-1}(u) S_{  \bar{v}}^{-1}P^{{  \tilde{v}v}(11) }_{\tilde{1}2}R^{
vv}_{23}(u)R^{  \tilde{v}v}_{\tilde{1}3}(u-\frac{1}{2})P^{{
\tilde{v}v}(11)}_{\tilde{1}2}S_{
\bar{v}}, \eea
where $S_{\bar{v}}$ is a $11\times 11$ diagonal constant matrix with the elements
\bea
S_{
\bar{v}}={\rm diag}\left[1,1,1,1,1,1,1,1,1,1,\sqrt{\frac{13}{61}}\right].\label{V-matrix}\eea
After taking the correspondence
\bea
|\phi^{(11)}_i\rangle\longrightarrow |\varphi^{(11)}_i\rangle,\quad i=1,\cdots,11,\label{Identification-2}
\eea
where the bases $\{|\phi^{(11)}_i\rangle\}$ and $\{|\varphi^{(11)}_i\rangle\}$ are given by (\ref{Basis-11}) and (\ref{Basis-11-1}) below, we
have
\bea
R_{\langle \tilde{1}2\rangle 3}^{\bar{v}v}(u)\equiv R_{\bar{1}\, 3}^{\bar{v}v}(u).\label{R11-R11}
\eea
Taking the correspondence
\bea
|\phi^{(16)}_1\rangle\longrightarrow|11\rangle^{(s)},\cdots,|\phi^{(16)}_4\rangle\longrightarrow|14\rangle^{(s)},|\phi^{(16)}_5\rangle\longrightarrow|21\rangle^{(s)},
\cdots, |\phi^{(16)}_{16}\rangle\longrightarrow|44\rangle^{(s)}, \label{Identification-3}
\eea
and after some calculations, we arrive at (\ref{Spinor}), namely,
\bea
&&R_{\langle 1234\rangle 5}(u)\equiv S_{12}R^{  sv}_{15}(u-\frac{1}{4})R^{
sv}_{25}(u-\frac{11}{4})S_{12}^{-1},\label{fu2asd-1111111}\eea
where $S_{12}$ is the $16\times 16$ gauge transformation on the tensor space $V^{(s)}\otimes V^{(s)}$  with the
non-zero matrix elements
\begin{eqnarray}
&&S^{11}_{12}=S^{12}_{13}=-S^{11}_{21}=-S^{14}_{24}=-S^{12}_{31}=S^{14}_{42}=S^{21}_{34}=-S^{21}_{43}=-a_4=-\sqrt{\frac{6}{11}},\nonumber\\
&&S^{13}_{14}=S^{13}_{23}=-S^{13}_{32}=-S^{13}_{41}=-b_4=-\sqrt{\frac{3}{11}},\nonumber\\
&&S^{22}_{11}=S^{24}_{42}=-S^{34}_{33}=S^{44}_{44}=c_4=\sqrt{2},\nonumber\\
&&S^{23}_{12}=S^{23}_{21}=-S^{32}_{13}=-S^{32}_{31}=S^{42}_{24}=S^{43}_{34}=S^{42}_{42}=S^{43}_{43}=d_4=1,\nonumber\\
&&S^{31}_{32}=-S^{31}_{14}=-S^{33}_{14}=S^{31}_{23}=-S^{33}_{23}=-S^{33}_{32}=-S^{31}_{41}=-S^{33}_{41}=e_4=\frac{1}{\sqrt{2}},\nonumber\\
&&S^{41}_{14}=-S^{41}_{23}=S^{41}_{32}=-S^{41}_{41}=g_4=\frac{1}{{2}}\sqrt{\frac{65}{22}}.
 \label{RsV-element}
\end{eqnarray}
With the help of the $4$-dimensional projector $P^{{  sv}(4) }_{21}$ given by (\ref{a5}) and after taking
the correspondence
\bea
|\psi'_i\rangle\longrightarrow|i\rangle^{(s)},\quad i=1,\cdots,4,  \label{Identification-4}
\eea
where the basis $\{|\psi'_i\rangle\}$ is given by (\ref{a51-1}) below,  we have
\bea && R^{sv}_{1\,3}(u)\equiv R_{\langle 12\rangle 3}^{
sv}(u) = \tilde{\rho}_0^{-1}(u+\frac{1}{4})
P^{{  sv}(4) }_{21}R^{  vv}_{13}(u+\frac{1}{4})R^{
sv}_{23}(u-1)P^{{  sv}(4)
}_{21}. \label{R4-R4}
 \eea

\subsection*{A.2: Projectors}

Here we list the projectors used in this paper. $P^{{  vv}(1) }_{12}$ is the one-dimensional projector
\bea
P^{{  vv}(1) }_{12}=|\psi_0\rangle\langle\psi_0|, \quad  P^{{  vv}(1) }_{21}= P^{{  vv}(1) }_{12}, \label{a1}
\eea
where
\bea
|\psi_0\rangle=\frac{1}{\sqrt{5}}(|15\rangle+|24\rangle+|33\rangle+|42\rangle+|51\rangle).\no
\eea
$P_{12}$ is a 11-dimensional projector
\bea
P_{12}=\sum_{i=1}^{11} |{\phi}^{(11)}_i\rangle\langle{\phi}^{(11)}_i|,\quad P_{21} =P_{12},\label{a2}
\eea
where
\bea
&&|{\phi}^{(11)}_1\rangle=\frac{1}{\sqrt{2}}(|12\rangle-|21\rangle), \quad |{\phi}^{(11)}_2\rangle=\frac{1}{\sqrt{2}}(|13\rangle-|31\rangle),\nonumber\\[4pt]
&&|{\phi}^{(11)}_3\rangle=\frac{1}{\sqrt{2}}(|14\rangle-|41\rangle), \quad |{\phi}^{(11)}_4\rangle=\frac{1}{\sqrt{2}}(|15\rangle-|51\rangle),\nonumber\\[4pt]
&&|{\phi}^{(11)}_5\rangle=\frac{1}{\sqrt{2}}(|23\rangle-|32\rangle),\quad |{\phi}^{(11)}_6\rangle=\frac{1}{\sqrt{2}}(|24\rangle-|42\rangle),\nonumber\\[4pt]
&&|{\phi}^{(11)}_7\rangle=\frac{1}{\sqrt{2}}(|25\rangle-|52\rangle),\quad |{\phi}^{(11)}_8\rangle=\frac{1}{\sqrt{2}}(|43\rangle-|34\rangle),\nonumber\\[4pt]
&&|{\phi}^{(11)}_{9}\rangle=\frac{1}{\sqrt{2}}(|53\rangle-|35\rangle),\quad |{\phi}^{(11)}_{10}\rangle=\frac{1}{\sqrt{2}}(|45\rangle-|54\rangle),\nonumber\\[4pt]
&&|{\phi}^{(11)}_{11}\rangle=\frac{1}{\sqrt{5}}(|15\rangle+|24\rangle+|33\rangle+|42\rangle+|51\rangle).\label{Basis-11}
\eea
$P_{123}$ is a 15-dimensional projector
\bea
P=\sum_{i=1}^{15}
|{\phi}^{(15)}_i\rangle\langle{\phi}^{(15)}_i|,\quad P_{321}=P_{123},\label{a3} \eea
where
\begin{eqnarray}
&&|{\phi}^{(15)}_1\rangle=\frac{1}{\sqrt{6}}(|123\rangle-|132\rangle-|213\rangle+|231\rangle+|312\rangle-|321\rangle),\nonumber\\[4pt]
&&|{\phi}^{(15)}_2\rangle=\frac{1}{\sqrt{6}}(|124\rangle-|142\rangle-|214\rangle+|241\rangle+|412\rangle-|421\rangle),\nonumber\\[4pt]
&&|{\phi}^{(15)}_3\rangle=\frac{1}{\sqrt{6}}(|125\rangle-|152\rangle-|215\rangle+|251\rangle+|512\rangle-|521\rangle),\nonumber\\[4pt]
&&|{\phi}^{(15)}_4\rangle=\frac{1}{\sqrt{6}}(|134\rangle-|143\rangle-|314\rangle+|341\rangle+|413\rangle-|431\rangle),\nonumber\\[4pt]
&&|{\phi}^{(15)}_5\rangle=\frac{1}{\sqrt{6}}(|135\rangle-|153\rangle-|315\rangle+|351\rangle+|513\rangle-|531\rangle),\nonumber\\[4pt]
&&|{\phi}^{(15)}_6\rangle=\frac{1}{\sqrt{6}}(|145\rangle-|154\rangle-|415\rangle+|451\rangle+|514\rangle-|541\rangle),\nonumber\\[4pt]
&&|{\phi}^{(15)}_7\rangle=\frac{1}{\sqrt{6}}(|234\rangle-|243\rangle-|324\rangle+|342\rangle+|423\rangle-|432\rangle),\nonumber\\[4pt]
&&|{\phi}^{(15)}_8\rangle=\frac{1}{\sqrt{6}}(|235\rangle-|253\rangle-|325\rangle+|352\rangle+|523\rangle-|532\rangle),\nonumber\\[4pt]
&&|{\phi}^{(15)}_9\rangle=\frac{1}{\sqrt{6}}(|245\rangle-|254\rangle-|425\rangle+|452\rangle+|524\rangle-|542\rangle),\nonumber\\[4pt]
&&|{\phi}^{(15)}_{10}\rangle=\frac{1}{\sqrt{6}}(|345\rangle-|354\rangle-|435\rangle+|453\rangle+|534\rangle-|543\rangle),\nonumber\\[4pt]
&&|{\phi}^{(15)}_{11}\rangle=\frac{1}{\sqrt{13}}(2|151\rangle+|124\rangle+|133\rangle+|142\rangle-|214\rangle+|241\rangle\nonumber\\[4pt]
&&\hspace{18mm}-|313\rangle+|331\rangle-|412\rangle+|421\rangle),\nonumber\\[4pt]
&&|{\phi}^{(15)}_{12}\rangle=\frac{1}{\sqrt{13}}(|215\rangle+2|242\rangle+|233\rangle+|251\rangle-|125\rangle+|152\rangle\nonumber\\[4pt]
&&\hspace{18mm}-|323\rangle+|332\rangle+|512\rangle-|521\rangle),\nonumber\\[4pt]
&&|{\phi}^{(15)}_{13}\rangle=\frac{1}{\sqrt{13}}(|513\rangle-|531\rangle+|423\rangle-|432\rangle+|315\rangle+|324\rangle\nonumber\\[4pt]
&&\hspace{18mm}+|333\rangle+|342\rangle+|351\rangle-|234\rangle+|243\rangle-|135\rangle+|153\rangle),\nonumber\\[4pt]
&&|{\phi}^{(15)}_{14}\rangle=\frac{1}{\sqrt{13}}(|415\rangle+2|424\rangle+|433\rangle+|451\rangle-|145\rangle+|154\rangle\nonumber\\[4pt]
&&\hspace{18mm}-|343\rangle+|334\rangle+|514\rangle-|541\rangle),\nonumber\\[4pt]
&&|{\phi}^{(15)}_{15}\rangle=\frac{1}{\sqrt{13}}(2|515\rangle+|524\rangle+|533\rangle+|542\rangle-|254\rangle+|245\rangle\nonumber\\[4pt]
&&\hspace{18mm}-|353\rangle+|335\rangle-|452\rangle+|425\rangle).\nonumber
\end{eqnarray}
$P_{1234}$ is a 16-dimensional projector
\bea
P_{1234}=\sum_{i=1}^{16}
|{\phi}^{(16)}_i\rangle\langle{\phi}^{(16)}_i|, \quad P_{4321}=P_{1234},\label{a4}
\eea where
\bea
&&|{\phi}^{(16)}_1\rangle=\frac{1}{2\sqrt{6}}[|4123\rangle-|4132\rangle-|4213\rangle+|4231\rangle+|4312\rangle-|4321\rangle\nonumber\\
&&\hspace{15mm}-|1423\rangle+|1432\rangle+|2413\rangle-|2431\rangle-|3412\rangle+|3421\rangle\nonumber\\
&&\hspace{15mm}+|1243\rangle-|1342\rangle-|2143\rangle+|2341\rangle+|3142\rangle-|3241\rangle\nonumber\\
&&\hspace{15mm}-|1234\rangle+|1324\rangle+|2134\rangle-|2314\rangle-|3124\rangle+|3214\rangle],\nonumber\\
&&|{\phi}^{(16)}_2\rangle=\frac{1}{2\sqrt{6}}[|5123\rangle-|5132\rangle-|5213\rangle+|5231\rangle+|5312\rangle-|5321\rangle\nonumber\\
&&\hspace{15mm}-|1523\rangle+|1532\rangle+|2513\rangle-|2531\rangle-|3512\rangle+|3521\rangle\nonumber\\
&&\hspace{15mm}+|1253\rangle-|1352\rangle-|2153\rangle+|2351\rangle+|3152\rangle-|3251\rangle\nonumber\\
&&\hspace{15mm}-|1235\rangle+|1325\rangle+|2135\rangle-|2315\rangle-|3125\rangle+|3215\rangle],\nonumber\\
&&|{\phi}^{(16)}_3\rangle=\frac{1}{2\sqrt{6}}[|5124\rangle-|5142\rangle-|5214\rangle+|5241\rangle+|5412\rangle-|5421\rangle\nonumber\\
&&\hspace{15mm}-|1524\rangle+|1542\rangle+|2514\rangle-|2541\rangle-|4512\rangle+|4521\rangle\nonumber\\
&&\hspace{15mm}+|1254\rangle-|1452\rangle-|2154\rangle+|2451\rangle+|4152\rangle-|4251\rangle\nonumber\\
&&\hspace{15mm}-|1245\rangle+|1425\rangle+|2145\rangle-|2415\rangle-|4125\rangle+|4215\rangle],\nonumber\\
&&|{\phi}^{(16)}_4\rangle=\frac{1}{2\sqrt{6}}[|5134\rangle-|5143\rangle-|5314\rangle+|5341\rangle+|5413\rangle-|5431\rangle\nonumber\\
&&\hspace{15mm}-|1534\rangle+|1543\rangle+|3514\rangle-|3541\rangle-|4513\rangle+|4531\rangle\nonumber\\
&&\hspace{15mm}+|1354\rangle-|1453\rangle-|3154\rangle+|3451\rangle+|4153\rangle-|4351\rangle\nonumber\\
&&\hspace{15mm}-|1345\rangle+|1435\rangle+|3145\rangle-|3415\rangle-|4135\rangle+|4315\rangle],\nonumber\\
&&|{\phi}^{(16)}_5\rangle=\frac{1}{2\sqrt{6}}[|5234\rangle-|5243\rangle-|5324\rangle+|5342\rangle+|5423\rangle-|5432\rangle\nonumber\\
&&\hspace{15mm}-|2534\rangle+|2543\rangle+|3524\rangle-|3542\rangle-|4523\rangle+|4532\rangle\nonumber\\
&&\hspace{15mm}+|2354\rangle-|2453\rangle-|3254\rangle+|3452\rangle+|4253\rangle-|4352\rangle\nonumber\\
&&\hspace{15mm}-|2345\rangle+|2435\rangle+|3245\rangle-|3425\rangle-|4235\rangle+|4325\rangle],\nonumber\\
&&|{\phi}^{(16)}_6\rangle=\frac{1}{2\sqrt{11}}[2|1251\rangle+2|1242\rangle+|1233\rangle-|1323\rangle+|1332\rangle\nonumber\\
&&\hspace{15mm}-2|2151\rangle-2|2142\rangle-|2133\rangle+|3123\rangle-|3132\rangle\nonumber\\
&&\hspace{15mm}+2|1512\rangle+2|2412\rangle+|2313\rangle-|3213\rangle+|3312\rangle\nonumber\\
&&\hspace{15mm}-2|1521\rangle-2|2421\rangle-|2331\rangle+|3231\rangle-|3321\rangle],\nonumber\\
&&|{\phi}^{(16)}_7\rangle=\frac{1}{2\sqrt{11}}[2|3151\rangle+|3142\rangle+|3133\rangle+|3124\rangle-2|1351\rangle-|1342\rangle\nonumber\\
&&\hspace{15mm}-|1333\rangle-|1324\rangle+2|1531\rangle+|1432\rangle-|3313\rangle+|1234\rangle\nonumber\\
&&\hspace{15mm}-2|1513\rangle-|1423\rangle+|3331\rangle-|1243\rangle-|3214\rangle+|3241\rangle\nonumber\\
&&\hspace{15mm}-|3412\rangle+|3421\rangle+|2314\rangle-|2341\rangle+|4312\rangle-|4321\rangle\nonumber\\
&&\hspace{15mm}-|2134\rangle+|2431\rangle-|4132\rangle+|4231\rangle\nonumber\\
&&\hspace{15mm}+|2143\rangle-|2413\rangle+|4123\rangle-|4213\rangle],\nonumber\\
&&|{\phi}^{(16)}_8\rangle=\frac{1}{2\sqrt{11}}[2|4151\rangle+2|4124\rangle+|4133\rangle-|4313\rangle+|4331\rangle\nonumber\\
&&\hspace{15mm}-2|1451\rangle-2|1424\rangle-|1433\rangle+|3413\rangle-|3431\rangle\nonumber\\
&&\hspace{15mm}+2|1541\rangle+2|4241\rangle+|1343\rangle-|3143\rangle+|3341\rangle\nonumber\\
&&\hspace{15mm}-2|1514\rangle-2|4214\rangle-|1334\rangle+|3134\rangle-|3314\rangle],\nonumber\\
&&|{\phi}^{(16)}_9\rangle=\frac{1}{2\sqrt{11}}[|5142\rangle+|5133\rangle+|5124\rangle-|5214\rangle+|5241\rangle-|1542\rangle\nonumber\\
&&\hspace{15mm}-|1533\rangle-|1524\rangle+|2514\rangle-|2541\rangle+|1452\rangle+|1353\rangle\nonumber\\
&&\hspace{15mm}+|1254\rangle-|2154\rangle+|2451\rangle-|1425\rangle-|1335\rangle-|1245\rangle\nonumber\\
&&\hspace{15mm}+|2145\rangle-|2415\rangle-|5313\rangle+|5331\rangle-|5412\rangle+|5421\rangle\nonumber\\
&&\hspace{15mm}+2|5151\rangle+|3513\rangle-|3531\rangle+|4512\rangle-|4521\rangle-2|1515\rangle\nonumber\\
&&\hspace{15mm}-|3153\rangle+|3351\rangle-|4152\rangle+|4251\rangle\nonumber\\
&&\hspace{15mm}+|3135\rangle-|3315\rangle+|4125\rangle-|4215\rangle],\nonumber\\
&&|{\phi}^{(16)}_{10}\rangle=\frac{1}{2\sqrt{11}}[2|3242\rangle+|3233\rangle+|3215\rangle+|3251\rangle\nonumber\\
&&\hspace{15mm}-2|2342\rangle-|2333\rangle-|2315\rangle-|2351\rangle+2|2432\rangle+|3332\rangle\nonumber\\
&&\hspace{15mm}+|2135\rangle+|2531\rangle-2|2423\rangle-|3323\rangle-|2153\rangle-|2513\rangle\nonumber\\
&&\hspace{15mm}-|3521\rangle+|3512\rangle-|3125\rangle+|3152\rangle+|5321\rangle-|5312\rangle\nonumber\\
&&\hspace{15mm}+|1325\rangle-|1352\rangle-|5231\rangle+|5132\rangle-|1235\rangle+|1532\rangle\nonumber\\
&&\hspace{15mm}+|5213\rangle-|5123\rangle+|1253\rangle-|1523\rangle],\nonumber\\
&&|{\phi}^{(16)}_{11}\rangle=\frac{1}{2\sqrt{11}}[|4215\rangle+|4233\rangle+|4251\rangle-|4323\rangle+|4332\rangle-|2415\rangle\nonumber\\
&&\hspace{15mm}-|2433\rangle-|2451\rangle+|3423\rangle-|3432\rangle+|2145\rangle+|2343\rangle\nonumber\\
&&\hspace{15mm}+|2541\rangle-|3243\rangle+|3342\rangle-|2154\rangle-|2334\rangle-|2514\rangle\nonumber\\
&&\hspace{15mm}+|3234\rangle-|3324\rangle-|4521\rangle+|4512\rangle-|4125\rangle+|4152\rangle\nonumber\\
&&\hspace{15mm}+2|4242\rangle+|5421\rangle-|5412\rangle+|1425\rangle-|1452\rangle-2|2424\rangle\nonumber\\
&&\hspace{15mm}-|5241\rangle+|5142\rangle-|1245\rangle+|1542\rangle\nonumber\\
&&\hspace{15mm}+|5214\rangle-|5124\rangle+|1254\rangle-|1524\rangle],\nonumber\\
&&|{\phi}^{(16)}_{12}\rangle=\frac{1}{2\sqrt{11}}[2|5215\rangle+2|5242\rangle+|5233\rangle-|5323\rangle+|5332\rangle\nonumber\\
&&\hspace{15mm}-2|2515\rangle-2|2542\rangle-|2533\rangle+|3523\rangle-|3532\rangle\nonumber\\
&&\hspace{15mm}+2|5152\rangle+2|2452\rangle+|2353\rangle-|3253\rangle+|3352\rangle\nonumber\\
&&\hspace{15mm}-2|5125\rangle-2|2425\rangle-|2335\rangle+|3235\rangle-|3325\rangle],\nonumber\\
&&|{\phi}^{(16)}_{13}\rangle=\frac{1}{\sqrt{65}}[2|5151\rangle+|5142\rangle+|5133\rangle+|5124\rangle-|5214\rangle+|5241\rangle\nonumber\\
&&\hspace{15mm}-|5313\rangle+|5331\rangle-|5412\rangle+|5421\rangle+|4215\rangle+|4233\rangle\nonumber\\
&&\hspace{15mm}+2|4242\rangle+|4251\rangle-|4323\rangle+|4332\rangle+|4512\rangle-|4521\rangle\nonumber\\
&&\hspace{15mm}+|4152\rangle-|4125\rangle+|3315\rangle+|3324\rangle+|3333\rangle+|3342\rangle\nonumber\\
&&\hspace{15mm}+|3351\rangle-|3135\rangle+|3153\rangle-|3234\rangle+|3243\rangle+|3423\rangle\nonumber\\
&&\hspace{15mm}-|3432\rangle-|3531\rangle+|3513\rangle+|2415\rangle+|2451\rangle+2|2424\rangle\nonumber\\
&&\hspace{15mm}+|2433\rangle-|2541\rangle+|2514\rangle-|2343\rangle+|2334\rangle+|2154\rangle\nonumber\\
&&\hspace{15mm}-|2145\rangle+2|1515\rangle+|1524\rangle+|1533\rangle+|1542\rangle+|1425\rangle\nonumber\\
&&\hspace{15mm}-|1452\rangle-|1353\rangle+|1335\rangle-|1254\rangle+|1245\rangle],\nonumber\\
&&|{\phi}^{(16)}_{14}\rangle=\frac{1}{2\sqrt{11}}[2|4324\rangle+|4333\rangle+|4315\rangle+|4351\rangle-2|3424\rangle-|3433\rangle\nonumber\\
&&\hspace{15mm}-|3415\rangle-|3451\rangle+2|4243\rangle+|3343\rangle+|3145\rangle+|3541\rangle\nonumber\\
&&\hspace{15mm}-2|4234\rangle-|3334\rangle-|3154\rangle-|3514\rangle-|4135\rangle+|4153\rangle\nonumber\\
&&\hspace{15mm}-|4531\rangle+|4513\rangle+|1435\rangle-|1453\rangle+|5431\rangle-|5413\rangle,\nonumber\\
&&\hspace{15mm}-|1345\rangle+|1543\rangle-|5341\rangle+|5143\rangle\nonumber\\
&&\hspace{15mm}+|1354\rangle-|1534\rangle+|5314\rangle-|5134\rangle],\nonumber\\
&&|{\phi}^{(16)}_{15}\rangle=\frac{1}{2\sqrt{11}}[2|5315\rangle+|5333\rangle+|5324\rangle+|5342\rangle-2|3515\rangle-|3533\rangle\nonumber\\
&&\hspace{15mm}-|3524\rangle-|3542\rangle+2|5153\rangle+|3353\rangle+|3254\rangle+|3452\rangle\nonumber\\
&&\hspace{15mm}-2|5135\rangle-|3335\rangle-|3245\rangle-|3425\rangle-|5234\rangle+|5243\rangle\nonumber\\
&&\hspace{15mm}-|5432\rangle+|5423\rangle+|2534\rangle-|2543\rangle+|4532\rangle-|4523\rangle\nonumber\\
&&\hspace{15mm}-|2354\rangle+|2453\rangle-|4352\rangle+|4253\rangle\nonumber\\
&&\hspace{15mm}+|2345\rangle-|2435\rangle+|4325\rangle-|4235\rangle],\nonumber\\
&&|{\phi}^{(16)}_{16}\rangle=\frac{1}{2\sqrt{11}}[2|5415\rangle+2|5424\rangle+|5433\rangle-|5343\rangle+|5334\rangle\nonumber\\
&&\hspace{15mm}-2|4515\rangle-2|4524\rangle-|4533\rangle+|3543\rangle-|3534\rangle\nonumber\\
&&\hspace{15mm}+2|5154\rangle+2|4254\rangle+|4353\rangle-|3453\rangle+|3354\rangle\nonumber\\
&&\hspace{15mm}-2|5145\rangle-2|4245\rangle-|4335\rangle+|3435\rangle-|3345\rangle].\no
 \eea
$P^{{  sv}(4) }_{12}$ is a $4$-dimensional projector
\bea
P^{{  sv}(4) }_{12}=\sum_{i=1}^4 |\psi_i\rangle\langle\psi_i|,\label{a5}
\eea
where
\bea
&&|\psi_1\rangle=\frac{1}{\sqrt{5}}(|1,3\rangle+\sqrt{2}|2,2\rangle+\sqrt{2}|3,1\rangle),\nonumber\\
&&|\psi_2\rangle=\frac{1}{\sqrt{5}}(\sqrt{2}|1,4\rangle-|2,3\rangle+\sqrt{2}|4,1\rangle),\nonumber\\
&&|\psi_3\rangle=\frac{1}{\sqrt{5}}(\sqrt{2}|1,5\rangle-|3,3\rangle-\sqrt{2}|4,2\rangle),\nonumber\\
&&|\psi_4\rangle=\frac{1}{\sqrt{5}}(\sqrt{2}|2,5\rangle-\sqrt{2}|3,4\rangle+|4,3\rangle).\label{a51}
\eea
Similarly, we can construct the projector $P^{{vs}(4) }_{21}$
\bea
P^{{  vs}(4) }_{21}=\sum_{i=1}^4 |\psi'_i\rangle\langle\psi'_i|,\label{a5-1}
\eea
where
\bea
&&|\psi'_1\rangle=\frac{1}{\sqrt{5}}(|3,1\rangle+\sqrt{2}|2,2\rangle+\sqrt{2}|1,3\rangle),\nonumber\\
&&|\psi'_2\rangle=\frac{1}{\sqrt{5}}(\sqrt{2}|4,1\rangle-|3,2\rangle+\sqrt{2}|1,4\rangle),\nonumber\\
&&|\psi'_3\rangle=\frac{1}{\sqrt{5}}(\sqrt{2}|5,1\rangle-|3,3\rangle-\sqrt{2}|2,4\rangle),\nonumber\\
&&|\psi'_4\rangle=\frac{1}{\sqrt{5}}(\sqrt{2}|5,2\rangle-\sqrt{2}|4,3\rangle+|3,4\rangle).\label{a51-1}
\eea
$P^{{  \bar{v}v}(5) }_{\bar{1}2}$ is the $5$-dimensional projector
\bea
P^{{  \bar{v}v}(5) }_{\bar{1}2}=\sum_{i=1}^5 |{\varphi}^{(5)}_i\rangle\langle{\varphi}^{(5)}_i|,\label{a6}
\eea
where
\bea  && |{\varphi}^{(5)}_1\rangle=\frac{1}{\sqrt{14}}(-|1,4\rangle-|2,3\rangle-|3,2\rangle-|4,1\rangle-\sqrt{10}|11,1\rangle),\nonumber\\
&&|{\varphi}^{(5)}_2\rangle=\frac{1}{\sqrt{14}}(|1,5\rangle-|5,3\rangle-|6,2\rangle-|7,1\rangle-\sqrt{10}|11,2\rangle),\nonumber\\
&&|{\varphi}^{(5)}_3\rangle=\frac{1}{\sqrt{14}}(|2,5\rangle+|5,4\rangle-|8,2\rangle-|91\rangle-\sqrt{10}|11,3\rangle),\nonumber\\
&&|{\varphi}^{(5)}_4\rangle=\frac{1}{\sqrt{14}}(|3,5\rangle+|6,4\rangle+|8,3\rangle-|10,1\rangle-\sqrt{10}|11,4\rangle),\nonumber\\
&&|{\varphi}^{(5)}_5\rangle=\frac{1}{\sqrt{14}}(|4,5\rangle+|7,4\rangle+|9,3\rangle+|10,2\rangle-\sqrt{10}|11,5\rangle).\label{Basis-5}
\eea
Because the dimension of $V_{\bar 1}$ and that of $V_{2}$ are not equal, we need introduce another $5$-dimensional projector
\bea
P^{{  v\bar{v}}(5) }_{2\bar{1}}=\sum_{i=1}^5 |{\varphi}^{'(5)}_i\rangle\langle{\varphi}^{'(5)}_i|,\label{a7}
\eea
where
\bea  && |{\varphi}^{'(5)}_1\rangle=\frac{1}{\sqrt{14}}(-|4,1\rangle-|3,2\rangle-|2,3\rangle-|1,4\rangle+\sqrt{10}|1,11\rangle),\nonumber\\
&&|{\varphi}^{'(5)}_2\rangle=\frac{1}{\sqrt{14}}(|5,1\rangle-|3,5\rangle-|2,6\rangle-|1,7\rangle+\sqrt{10}|2,11\rangle),\nonumber\\
&&|{\varphi}^{'(5)}_3\rangle=\frac{1}{\sqrt{14}}(|5,2\rangle+|4,5\rangle-|2,8\rangle-|1,9\rangle+\sqrt{10}|3,11\rangle),\nonumber\\
&&|{\varphi}^{'(5)}_4\rangle=\frac{1}{\sqrt{14}}(|5,3\rangle+|4,6\rangle+|3,8\rangle-|1,10\rangle+\sqrt{10}|4,11\rangle),\nonumber\\
&&|{\varphi}^{'(5)}_5\rangle=\frac{1}{\sqrt{14}}(|5,4\rangle+|4,7\rangle+|3,9\rangle+|2,10\rangle+\sqrt{10}|5,11\rangle). \no
\eea
$P^{{  \tilde{v}v}(11) }_{\tilde{1}2}$ is the 11-dimensional projector
\bea
P^{{  \tilde{v}v}(11) }_{\tilde{1}2}=\sum_{i=1}^{11} |{\varphi}^{(11)}_i\rangle\langle{\varphi}^{(11)}_i|,\label{a8}
\eea
where
\bea  && |{\varphi}^{(11)}_1\rangle=\frac{1}{\sqrt{61}}(-\sqrt{3}|1,3\rangle-\sqrt{3}|2,2\rangle-\sqrt{3}|3,1\rangle+\sqrt{26}|11,2\rangle-\sqrt{26}|12,1\rangle),\nonumber\\
&&|{\varphi}^{(11)}_2\rangle=\frac{1}{\sqrt{61}}(\sqrt{3}|1,4\rangle-\sqrt{3}|4,2\rangle-\sqrt{3}|5,1\rangle+\sqrt{26}|11,3\rangle-\sqrt{26}|13,1\rangle),\nonumber\\
&&|{\varphi}^{(11)}_3\rangle=\frac{1}{\sqrt{61}}(\sqrt{3}|2,4\rangle+\sqrt{3}|4,3\rangle-\sqrt{3}|6,1\rangle+\sqrt{26}|11,4\rangle-\sqrt{26}|14,1\rangle),\nonumber\\
&&|{\varphi}^{(11)}_4\rangle=\frac{1}{\sqrt{61}}(\sqrt{3}|3,4\rangle+\sqrt{3}|5,3\rangle+\sqrt{3}|6,2\rangle+\sqrt{26}|11,5\rangle-\sqrt{26}|15,1\rangle),\nonumber\\
&&|{\varphi}^{(11)}_5\rangle=\frac{1}{\sqrt{61}}(-\sqrt{3}|1,5\rangle-\sqrt{3}|7,2\rangle-\sqrt{3}|8,1\rangle+\sqrt{26}|12,3\rangle-\sqrt{26}|13,2\rangle),\nonumber\\
&&|{\varphi}^{(11)}_6\rangle=\frac{1}{\sqrt{61}}(-\sqrt{3}|2,5\rangle+\sqrt{3}|7,3\rangle-\sqrt{3}|9,1\rangle+\sqrt{26}|12,4\rangle-\sqrt{26}|14,2\rangle),\nonumber\\
&&|{\varphi}^{(11)}_7\rangle=\frac{1}{\sqrt{61}}(-\sqrt{3}|3,5\rangle+\sqrt{3}|8,3\rangle+\sqrt{3}|9,2\rangle+\sqrt{26}|12,5\rangle-\sqrt{26}|15,2\rangle),\nonumber\\
&&|{\varphi}^{(11)}_8\rangle=\frac{1}{\sqrt{61}}(-\sqrt{3}|4,5\rangle-\sqrt{3}|7,4\rangle-\sqrt{3}|10,1\rangle+\sqrt{26}|13,4\rangle-\sqrt{26}|14,3\rangle),\nonumber\\
&&|{\varphi}^{(11)}_9\rangle=\frac{1}{\sqrt{61}}(-\sqrt{3}|5,5\rangle-\sqrt{3}|8,4\rangle+\sqrt{3}|10,2\rangle+\sqrt{26}|13,5\rangle-\sqrt{26}|15,3\rangle),\nonumber\\
&&|{\varphi}^{(11)}_{10}\rangle=\frac{1}{\sqrt{61}}(-\sqrt{3}|6,5\rangle-\sqrt{3}|9,4\rangle-\sqrt{3}|10,3\rangle+\sqrt{26}|14,5\rangle-\sqrt{26}|15,4\rangle),\nonumber\\
&&|{\varphi}^{(11)}_{11}\rangle=\frac{1}{\sqrt{5}}(|11,5\rangle+|12,4\rangle+|13,3\rangle+|14,2\rangle+|15,1\rangle).\label{Basis-11-1}
\eea
Because the dimensions of spaces $V_{\tilde 1}$ and $V_2$ in the operator $P^{{  \tilde{v}v}(11) }_{\tilde{1}2}$ are not equal, we should introduce the related projector
\bea
P^{{  v\tilde{v}}(11) }_{2\tilde{1}}=\sum_{i=1}^{11} |{\varphi}^{'(11)}_i\rangle\langle{\varphi}^{'(11)}_i|,\label{a9}
\eea
where
\bea
&& |{\varphi}^{'(11)}_1\rangle=\frac{1}{\sqrt{61}}(-\sqrt{3}|3,1\rangle-\sqrt{3}|2,2\rangle-\sqrt{3}|1,3\rangle-\sqrt{26}|2,11\rangle+\sqrt{26}|1,12\rangle),\nonumber\\
&&|{\varphi}^{'(11)}_2\rangle=\frac{1}{\sqrt{61}}(\sqrt{3}|4,1\rangle-\sqrt{3}|2,4\rangle-\sqrt{3}|1,5\rangle-\sqrt{26}|3,11\rangle+\sqrt{26}|1,13\rangle),\nonumber\\
&&|{\varphi}^{'(11)}_3\rangle=\frac{1}{\sqrt{61}}(\sqrt{3}|4,2\rangle+\sqrt{3}|3,4\rangle-\sqrt{3}|1,6\rangle-\sqrt{26}|4,11\rangle+\sqrt{26}|1,14\rangle),\nonumber\\
&&|{\varphi}^{'(11)}_4\rangle=\frac{1}{\sqrt{61}}(\sqrt{3}|4,3\rangle+\sqrt{3}|3,5\rangle+\sqrt{3}|2,6\rangle-\sqrt{26}|5,11\rangle+\sqrt{26}|1,15\rangle),\nonumber\\
&&|{\varphi}^{'(11)}_5\rangle=\frac{1}{\sqrt{61}}(-\sqrt{3}|5,1\rangle-\sqrt{3}|2,7\rangle-\sqrt{3}|1,8\rangle-\sqrt{26}|3,12\rangle+\sqrt{26}|2,13\rangle),\nonumber\\
&&|{\varphi}^{'(11)}_6\rangle=\frac{1}{\sqrt{61}}(-\sqrt{3}|5,2\rangle+\sqrt{3}|3,7\rangle-\sqrt{3}|1,9\rangle-\sqrt{26}|4,12\rangle+\sqrt{26}|2,14\rangle),\nonumber\\
&&|{\varphi}^{'(11)}_7\rangle=\frac{1}{\sqrt{61}}(-\sqrt{3}|5,3\rangle+\sqrt{3}|3,8\rangle+\sqrt{3}|2,9\rangle-\sqrt{26}|5,12\rangle+\sqrt{26}|2,15\rangle),\nonumber\\
&&|{\varphi}^{'(11)}_8\rangle=\frac{1}{\sqrt{61}}(-\sqrt{3}|5,4\rangle-\sqrt{3}|4,7\rangle-\sqrt{3}|1,10\rangle-\sqrt{26}|4,13\rangle+\sqrt{26}|3,14\rangle),\nonumber\\
&&|{\varphi}^{'(11)}_9\rangle=\frac{1}{\sqrt{61}}(-\sqrt{3}|5,5\rangle-\sqrt{3}|4,8\rangle+\sqrt{3}|2,10\rangle-\sqrt{26}|5,13\rangle+\sqrt{26}|3,15\rangle),\nonumber\\
&&|{\varphi}^{'(11)}_{10}\rangle=\frac{1}{\sqrt{61}}(-\sqrt{3}|5,6\rangle-\sqrt{3}|4,9\rangle-\sqrt{3}|3,10\rangle-\sqrt{26}|5,14\rangle+\sqrt{26}|4,15\rangle),\nonumber\\
&&|{\varphi}^{'(11)}_{11}\rangle=\frac{1}{\sqrt{5}}(|5,11\rangle+|4,12\rangle+|3,13\rangle+|2,14\rangle+|1,15\rangle). \no
\eea

\section*{Appendix B: Fusion of the $K$-matrices}
\setcounter{equation}{0}
\renewcommand{\theequation}{B.\arabic{equation}}

\subsection*{B.1: Associated spinorial reflection matrices}

We  introduce the corresponding spinorial $K$-matrix
$K^{s-}(u)$ and the dual reflecting matrix $K^{   s+}(u)$, which satisfy the reflection equation
\begin{equation}
 R^{   ss}_{12}(u-v){K^{  s-}_{1}}(u)R^{  ss}_{21}(u+v) {K^{  s-}_{2}}(v)=
 {K^{   s-}_{2}}(v)R^{  ss}_{12}(u+v){K^{  s-}_{1}}(u)R^{  ss}_{21}(u-v),
 \label{rss1}
 \end{equation} and its dual
\begin{eqnarray}
 & &R^{   ss}_{12}(-u+v){K^{   s+}_{1}}(u)R^{  ss}_{21}
 (-u-v-3){K^{   s+}_{2}}(v)\nonumber\\
&&\qquad\qquad\qquad\qquad={K^{   s+}_{2}}(v)R^{   ss}_{12}(-u-v-3) {K^{
s+}_{1}}(u)R^{  ss}_{21}(-u+v),
 \label{rsssll}
 \end{eqnarray}
respectively. The matrix forms of $K^{s\pm}(u)$ are
\bea
&& K^{  s-}(u)=M^{  s}, \quad
M^{  s}= \left(\begin{array}{cccc}1 &0&-c_{1}&0\\[6pt]
0&1 &0&c_{1}\\[6pt]
-c_{2}&0 &-1 &0\\[6pt]
0&c_{2}&0&-1 \end{array}\right), \label{K-matwwwrix-VV} \\[8pt]
&& K^{  s+}(u)=\tilde{M}^{  s}, \quad
\tilde{M}^{  s}=\left(\begin{array}{cccc}1 &0&-\tilde{c}_{1}&0\\[6pt]
0&1 &0&\tilde{c}_{1}\\[6pt]
-\tilde{c}_{2}&0 &-1 &0\\[6pt]
0&\tilde{c}_{2}&0&-1 \end{array}\right). \label{K-matwwwrix-wwVV} \eea
With the help of the
projector $P_{12}^{{  ss}(5)}$ given by (\ref{Project-v}), one can construct the fused
$K$-matrices \cite{Hao14} as follows:
\bea
&& K_{\langle 12\rangle }^{  v-}(u)=(u+\frac{3}{4})^{-1}
P_{12}^{{  ss}(5)}K_{2}^{  s-}(u+\frac{1}{4})R_{12}^{  ss}(2u)K_{1}^{  s-}(u-\frac{1}{4})P_{21}^{{  ss}(5)},\\
&&K_{\langle 12\rangle }^{  v+}(u) =-(u+\frac{3}{4})^{-1} P_{12}^{{  ss}(5)}K_{2}^{  s+}(u-\frac{1}{4})R_{12}^{  ss}(-2u-3)K_{1}^{  s+}(u+\frac{1}{4})P_{21}^{{  ss}(5)}.
\eea
We remark that $P_{21}^{{  ss}(5)}=P_{12}^{{ss}(5)}$. With the help of the equivalence (\ref{Identification}), we have
\bea
K^{v\pm}_1(u)\equiv  K_{\langle 12\rangle }^{  v\pm}(u).
\eea

\subsection*{B.2: Other fused $K$-matrices}

The one-dimensional projector $P_{21}^{  vv(1)}$ given by (\ref{Int-R1}) allows us to compute the  quantum determinants ${\rm Det}_q(K^{v\pm}(u))$
\bea && P_{21}^{  vv(1)}K_{1}^{  v-}(u)R_{21}^{  vv}(2u-\frac{3}{{2}})K_{2}^{  v-}(u-\frac{3}{{2}})P_{12}^{  vv(1)}\stackrel{{\rm def}}{=}{\rm Det}_q(K^{v-}(u))\,P_{12}^{  vv(1)},\label{fk-1}\\
&& P_{12}^{  vv(1)}K_{2}^{  v+}(u-\frac{3}{{2}})R_{12}^{  vv}(-2u-\frac{3}{{2}})K_{1}^{  v+}(u)P_{21}^{  vv(1)}\stackrel{{\rm def}}{=}
{\rm Det}_q(K^{v+}(u))\,P_{21}^{vv(1)}, \label{fk-2}\eea where
\bea
&&\hspace{-1.2truecm}{\rm Det}_q(K^{v-}(u))=-2^2(u-\frac{3}{{2}})(u-\frac{1}{{4}})h(u)h(-u),\quad h(u)=(1+c_1c_2)(4u+1),\no\\
&&\hspace{-1.2truecm}{\rm Det}_q(K^{v+}(u))
=-2^2(u+\frac{3}{{2}})(u+\frac{1}{{4}})\tilde{h}(u)\tilde{h}(-u),\quad
\tilde{h}(u)=(1+\tilde{c}_1\tilde{c}_2)(4u+1).
\eea
Using the the $11$-dimensional projector $P_{21}^{  vv}$ given by (\ref{a2}), we can construct the  $11\times 11$ $K$-matrices $K^{  \bar{v} \pm}(u)$
\bea
&&\hspace{-1.42truecm}K^{  \bar{v}-}_{\bar{1}}(u)\equiv\bar K_{\langle12\rangle}^-(u\hspace{-0.12truecm}+\hspace{-0.12truecm}\frac{1}{2})=\frac{1}{2(u-\frac{1}{2})h(u+\frac{1}{2})}\,
P_{21}K_{1}^{  v-}(u+\frac{1}{2})R_{21}^{  vv}(2u)K_{2}^{  v-}(u-\frac{1}{2})P_{12},\label{K-matrix-Bar-}\\
&&\hspace{-1.42truecm}K^{  \bar{v}+}_{\bar{1}}(u)\equiv\bar
K_{\langle12\rangle}^+(u\hspace{-0.12truecm}+\hspace{-0.12truecm}\frac{1}{2})
\hspace{-0.12truecm}=\hspace{-0.12truecm}
\frac{1}{2(u\hspace{-0.12truecm}+\hspace{-0.12truecm}2)\tilde{h}(u+\frac{1}{2})}\,P_{12}K_{2}^{
v+}(u\hspace{-0.12truecm}-\hspace{-0.12truecm}\frac{1}{2})
R_{12}^{  vv}(-2u-3)K_{1}^{
v+}(u\hspace{-0.12truecm}+\hspace{-0.12truecm}\frac{1}{2})P_{21}.\label{K-matrix-Bar+}
\eea The 15-dimensional projector $P_{321}$ given by (\ref{a3})
allows us to construct the $15\times 15$ $K$-matrices
$K^{\tilde{v}\pm}(u)$ \bea
\hspace{-1.2truecm}K^{\tilde{v}-}(u)\equiv \bar
K_{\langle123\rangle}^-(u+1)
&=&\hspace{-0.2truecm}\frac{P_{321}K_{1}^{
v-}(u+1)R_{21}^{  vv}(2u+1)R_{31}^{  vv}(2u)}{2^5(u+\frac{5}{4})(u+\frac{3}{4})(u-\frac{1}{2})u(u-1)h(u+1)h(u)}\nonumber\\[4pt]
&&\quad\times K^{  v-}_{2}(u)R_{32}^{  vv}(2u-1)K_{3}^{  v-}(u-1)P_{123},\label{K-matrix-tilde-}\\[4pt]
\hspace{-1.2truecm}K^{\tilde{v}+}(u)\equiv\bar K_{\langle123\rangle}^+(u+1)&=&\hspace{-0.2truecm}\frac{-P_{123}K_{3}^{
v+}(u-1)R_{23}^{  vv}(-2u-2)R_{13}^{  vv}(-2u-3)}{2^5(u+\frac{1}{4})(u+\frac{3}{4})(u+\frac{3}{2})(u+\frac{5}{2})(u+2)\tilde{h}(u+1)\tilde{h}(u)}\nonumber\\[4pt]
&&\quad\times K^{  v+}_{2}(u)R_{12}^{  vv}(-2u-4)
K_{1}^{  v+}(u+1)P_{321}.\label{K-matrix-tilde+}
\eea
All these $K$-matrices satisfy the associate reflection equations or the dual reflection equations.

Using the $5$-dimensional projectors $P_{\bar{1}2}^{  \bar{v}v(5)}$ and $P_{2\bar{1}}^{  \bar{v}v(5)}$ given by (\ref{a7})-(\ref{a8}) and
the correspondence (\ref{Identification-1}), we have
\bea
\hspace{-1.2truecm} K^{v-}_1(u)&\equiv& K^{  v-}_{\langle \bar{1}2\rangle }(u+\frac{1}{2})=- \frac{P_{\bar{1}2}^{  \bar{v}v(5)}K_{2}^{  v-}(u+\frac{1}{2})
R_{12}^{  \bar{v}v}(2u)K_{\bar{1}}^{ \bar{v}-}(u-\frac{1}{2})P_{2\bar{1}}^{  v\bar{v}(5)}}
{2^3(u-\frac{1}{2})(u+\frac{3}{4})(u+\frac{1}{4})h(u+\frac{1}{2})},\\[4pt]
\hspace{-1.2truecm}K^{v+}_1(u)&\equiv&K^{v+}_{\langle
\bar{1}2\rangle }(u+\frac{1}{2})=- \frac{P_{2\bar{1}}^{
v\bar{v}(5)}K_{ \bar{1}}^{ \bar{v}+}(u-\frac{1}{2})R_{2\bar{1}}^{
v\bar{v}}(-2u-3)K_{2}^{  v+}(u+\frac{1}{2})P_{\bar{1}2}^{
\bar{v}v(5)}}
{2^3(u+2)(u+\frac{1}{4})(u+\frac{3}{4})\tilde{h}(u+\frac{1}{2})}.\eea
Similarly with the help of the $11$-dimensional projectors
$P_{\tilde{1}2}^{  \tilde{v}v(11)}$ and $P_{2\tilde{1}}^{
\tilde{v}v(11)}$ given by (\ref{a8})-(\ref{a9}) and the
correspondence (\ref{Identification-2}), we have \bea
\hspace{-1.2truecm}K^{\bar{v}-}_{\bar{1}}(u)&\equiv &S_{
\bar{v}}^{-1}K^{ \bar{v}-}_{\langle \tilde{1}2\rangle }(u)S_{
\bar{v}}= \frac{S_{ \bar{v}}^{-1}P_{\tilde{1}2}^{
\tilde{v}v(11)}K_{2}^{  v-}(u)R^{
\tilde{v}v}_{\tilde{1}2}(2u-\frac{1}{2}) K_{\tilde{1}}^{
\tilde{v}-}(u-\frac{1}{2})P_{2\tilde{1}}^{  v\tilde{v}}S_{
\bar{v}}}
{2^2(u-\frac{1}{4})(u-\frac{1}{2})h(u)},\label{K-bar-}\\[4pt]
\hspace{-1.2truecm}K^{\bar{v}+}_{\bar{1}}(u)&\equiv & S_{
\bar{v}}^{-1}K^{ \bar{v}+}_{\langle \tilde{1}2\rangle }(u)S_{
\bar{v}}= -\frac{S_{ \bar{v}}^{-1}P_{2\tilde{1}}^{
v\tilde{v}(11)}K_{\tilde{1}}^{ \tilde{v}+}(u-\frac{1}{2})R^{
v\tilde{v}}_{2\tilde{1}}(-2u-\frac{5}{2}) K_{2}^{
v+}(u)P_{\tilde{1}2}^{  \tilde{v}v(11)}S_{ \bar{v}}}
{2^2(u+\frac{1}{4})(u+\frac{3}{2})\tilde{h}(u)},\label{K-bar+}\eea
where the  $11\times 11$ diagonal matrix $S_{\bar{v}}$ is given by
(\ref{V-matrix}). Moreover the $16$-dimensional projector $P_{4321}$ and the correspondence (\ref{Identification-3}) allow
us to have the identifications: \bea \bar
K_{\langle1234\rangle}^-(u)&=&P_{4321}K^{ v-}_1(u)R^{
vv}_{21}(2u-1)R^{  vv}_{31}(2u-2)R^{
vv}_{41}(2u-3)K^{  v-}_2(u-1)\no\\[4pt]
&&\times R^{  vv}_{32}(2u-3)R^{
vv}_{42}(2u-4)K^{  v-}_3(u-2)R^{
vv}_{43}(2u-5)K^{  v-}_4(u-3)P_{1234}\no\\[4pt]
&\equiv & \tilde{\rho}^-_2(u)S_{12}K^{
s-}_{1}(u-\frac{1}{4})R^{  ss}_{21}(2u-3)K^{
s-}_{2}(u-\frac{11}{4})S_{12}^{-1}, \\[4pt]
\bar  K_{\langle1234\rangle}^+(u)&=&P_{1234}K^{ v+}_4(u-3)R^{
vv}_{34}(-2u+2)R^{  vv}_{24}(-2u+1)R^{
vv}_{14}(-2u)K^{  v+}_3(u-2)\no\\[4pt]
&&\times R^{  vv}_{23}(-2u)R^{
vv}_{13}(-2u-1)K^{  v+}_2(u-1)R^{
vv}_{12}(-2u-2)K^{  v+}_1(u)P_{4321}\no\\[4pt]
&\equiv & \tilde{\rho}^+_2(u)S_{12}K^{
s+}_{2}(u-\frac{11}{4})R^{  ss}_{12}(-2u)K^{
s+}_{1}(u-\frac{1}{4})S_{12}^{-1}, \label{fk-14}\eea
where the  $16\times16$ constant matrix is given by (\ref{RsV-element}) and
\bea
\tilde{\rho}^-_2(u)&=&-2^{12}(u-1)(u-2)^2(u-3)(u-\frac{3}{{2}})
(u-\frac{5}{{2}})(u-\frac{3}{{4}})(u-\frac{1}{{4}})^2\no\\[4pt]
&&\times (u-\frac{5}{{4}})(u+\frac{1}{{4}}) h(u)h(u-1)h(u-2),\no\\[4pt]
\tilde{\rho}^+_2(u)&=&-2^{12}u(u+1)(u+\frac{1}{{2}})^2(u-\frac{1}{{2}})(u+\frac{3}{{2}})
(u-\frac{3}{{4}})(u-\frac{7}{{4}})\no\\[4pt]
&&\times (u-\frac{5}{{4}})^2(u-\frac{1}{{4}}) \tilde{h}(u)\tilde{h}(u-1)\tilde{h}(u-2).\no
\eea

In order to complete the whole fusion of  the $K$-matrices, we
need the fusion between vectorial $K_{1}^{  v-}(u)$ and spinorial
$K_{2}^{  s-}(u)$. The $4$-dimensional projectors
$P_{12}^{  sv(4)}$ and $P_{21}^{  sv(4)}$ given by
(\ref{a5})-(\ref{a5-1}) and the correspondence
(\ref{Identification-4}) enable us to have the identifications:
\bea
\hspace{-1.2truecm}K^{s-}(u)&\equiv &K_{\langle 12\rangle }^{
s-}(u+\frac{1}{{4}})=- \frac{P_{21}^{  sv(4)}K_{1}^{
v-}(u+\frac{1}{{4}})R_{21}^{  sv}(2u-\frac{3}{{4}})K_{2}^{
s-}(u-1)P_{12}^{  vs(4)}}
{2u\,h(u+\frac{1}{{4}})},\\[4pt]
\hspace{-1.2truecm}K^{s-}(u)&\equiv &K_{\langle 12\rangle }^{s+}(u+\frac{1}{{4}})=-
\frac{P_{12}^{  vs(4)}K_{2}^{  s+}(u-1)R_{12}^{  vs}(-2u-\frac{9}{{4}})K_{1}^{  v+}(u+\frac{1}{{4}})P_{21}^{  sv(4)}}
{2(u+\frac{7}{{2}})\tilde{h}(u+\frac{1}{{4}})}.\eea

\section*{Appendix C: Operator identities for the open case}
\setcounter{equation}{0}
\renewcommand{\theequation}{C.\arabic{equation}}
Similar to the periodic case (see (\ref{Fused-M-1})-(\ref{Fused-M-1})),  let us introduce some fused transfer matrices as follows:
\bea
\bar t_m (u)=tr_{12\cdots m} \{ \bar K_{\langle12\cdots m \rangle}^{{v}+}(u) \bar T_{\langle12\cdots m \rangle}^{ {v}}(u)
\bar K_{\langle12\cdots m \rangle}^{  {v}-}(u)\hat{\bar T}_{\langle12\cdots m \rangle}^{  {v}}(u)\}, \quad m=2,3,4,
\eea
where
\bea
&& \bar K_{\langle12 \cdots m\rangle}^-(u)=P_{m\cdots 1}
K^{v-}_1(u)R^{  vv}_{21}(2u-1)R^{  vv}_{31}(2u-2)\cdots \no \\[4pt]
&& \hspace{3cm}  \times  R^{
vv}_{m1}(2u-m+1) \bar K_{\langle2 \cdots m\rangle}^-(u-1) P_{m\cdots 1}, \no \\[4pt]
&&\bar K_{\langle12 \cdots m\rangle}^+(u)=P_{12\cdots m} \bar K_{\langle 2 \cdots m\rangle}^+(u-1)
R^{vv}_{1m}(-2u+(m-1)-3) \cdots \no \\[4pt]
&& \hspace{3cm}  \times  R^{vv}_{12}(-2u+1-3) K^{v+}_1(u)
  P_{12\cdots m}, \no\\[4pt]
&& \bar T_{\langle 12\cdots m\rangle}^{  v}(u)=P_{m\cdots 21}T_1^{
v}(u)T_2^{  v}(u-1) T_3^{  v}(u-2) \cdots T_m^{  v}(u-m+1)P_{m\cdots 21},\no \\[4pt]
&& \hat {\bar T}_{\langle 12\cdots m\rangle}^{  v}(u)=P_{12\cdots m}\hat T_1^{
v}(u)\hat T_2^{  v}(u-1) \hat T_3^{  v}(u-2) \cdots \hat T_m^{  v}(u-m+1)P_{12\cdots m}.\no
\eea

The fused reflecting monodromy matrices
satisfy the fusion relations (c.f., (\ref{Relation-1})-(\ref{Relation-3}) for the periodic case)
\bea
&&P^{{  vv}(1) }_{12}\hat{T}_1^{  v}(u)\hat{T}_2^{
v}(u-\frac{3}{{2}})P^{{  vv}(1) }_{12}=\prod_{i=1}^N
a_1(u+\theta_i)e_1(u+\theta_i-\frac{3}{{2}})P^{{  vv}(1)
}_{12}\times {\rm id},  \no \\[4pt]
&&\hat{\bar T}_{\langle12\rangle}^{  v}(u)=P_{12}\hat{T}_1^{
v}(u)\hat{T}_2^{  v}(u-1)P_{12}=\prod_{i=1}^N
\tilde{\rho}_0(u+\theta_i)\hat{T}_{\bar{1}}^{  \bar{v}}(u-\frac{1}{{2}}),\no \\[4pt]
&&\hat{\bar T}_{\langle123\rangle}^{  v}(u)=P_{123}\hat{T}_1^{
v}(u)\hat{T}_2^{  v}(u-1)\hat{T}_3^{  v}(u-2)P_{123}\no\\[4pt]
&&\hspace{20mm} =\prod_{i=1}^N
\tilde{\rho}_0(u+\theta_i)\tilde{\rho}_0(u+\theta_i-1)\hat{T}_{\tilde{1}}^{  \tilde{v}}(u-1),\no\\[4pt]
&&\hat{\bar T}_{\langle1234\rangle}^{  v}(u)=P_{1234}\hat{T}_1^{
v}(u)\hat{T}_2^{  v}(u-1)\hat{T}_3^{
v}(u-2)\hat{T}_4^{  v}(u-3)P_{1234}\no\\[4pt]
&&\hspace{20mm} =\prod_{i=1}^N \tilde{\rho}_1
(u+\theta_i)S_{12}\hat{T}_{1}^{  s}(u-\frac{1}{{4}})\hat{T}_{2}^{  s}(u-\frac{11}{{4}})S_{12}^{-1},\no \\[4pt]
&&P^{{  v\bar{v}}(5) }_{2\bar{1}}\hat{T}_2^{
v}(u)\hat{T}_{\bar{1}}^{  \bar{v}}(u-1)P^{{  v\bar{v}}(5)
}_{2\bar{1}}=\prod_{i=1}^N
\tilde{\rho}_0(u+\theta_i)\hat{T}_{\langle \bar{1}2\rangle }^{
{v}}(u-\frac{1}{{2}}),\no\\[4pt]
&&P^{{  v\tilde{v}}(11) }_{2\tilde{1}}\hat{T}_2^{
v}(u)\hat{T}_{\tilde{1}}^{  \tilde{v}}(u-\frac{1}{{2}})P^{{
v\tilde{v}}(11) }_{2\tilde{1}}=\prod_{i=1}^N
\tilde{\rho}_0(u+\theta_i)S_{  \bar{v}
}\hat{T}_{\langle \tilde{1}2\rangle }^{
\bar{v}}(u)S_{  \bar{v} }^{-1},\no\\[4pt]
&&P^{{  vs}(4) }_{21}\hat{T}_2^{  v}(u)\hat{T}_{{1}}^{
s}(u-\frac{5}{{4}})P^{{  vs}(4) }_{21}=\prod_{i=1}^N
\tilde{\rho}_0(u+\theta_i)\hat{T}_{\langle {1}2\rangle }^{
s}(u-\frac{1}{{4}}).\label{futt-7}
\eea
The above relations imply that  the fused transfer matrices are indeed proportional to those given by (\ref{2c0-1}) by some polynomials
(c.f., (\ref{bet1}) for the periodic case)
\bea
&&\bar t_2(u)={2^{6}}(u-1)(u+\frac{3}{{2}})(u+\frac{1}{{4}})^2H(u)t_2(u-\frac{1}{{2}}),  \\
&&\bar t_3(u)=-2^{18}(u+\frac{1}{{4}})^3(u-\frac{1}{{4}})^2(u-\frac{3}{{2}})(u-1)(u-2)(u-\frac{3}{{4}})^3\nonumber\\
&&\hspace{15mm}\times
(u+\frac{1}{{2}})(u+\frac{3}{{2}})(u+1) H(u)H(u-1)
t_3(u-1),  \\
&&\bar t_4(u)=2^{36}(u-1)(u-2)^2(u-3)(u-\frac{3}{{2}})(u-\frac{5}{{2}})(u-\frac{3}{{4}})^4(u-\frac{5}{{4}})^3
\nonumber\\
&&\hspace{15mm}\times(u-\frac{1}{{4}})^3(u+\frac{1}{{4}})^3u(u+1)(u+\frac{1}{{2}})^2(u-\frac{1}{{2}})(u+\frac{3}{{2}})(u-\frac{7}{{4}})^3
\nonumber\\
&&\hspace{15mm}\times \rho_2(2u-\frac32)\times H(u)H(u-1) H(u-2) t_{
s}(u-\frac{1}{{4}})t_{  s}(u-\frac{11}{{4}}). \label{bet} \eea

Similar to  (\ref{fui-7}) for the periodic case, we can derive the relations among the reflecting monodromy matrices given by (\ref{Mon-2})
\bea &&\hat{T}_1^{  v}(-\theta_j)\hat{T}_2^{
v}(-\theta_j-\frac{3}{{2}})=P^{{  vv}(1) }_{12}\hat{T}_1^{
v}(-\theta_j)\hat{T}_2^{  v}(-\theta_j-\frac{3}{{2}}),\no\\
&&\hat{T}_1^{  v}(-\theta_j)\hat{T}_2^{
v}(-\theta_j-1)=P_{12}\hat{T}_1^{
v}(-\theta_j)\hat{T}_2^{  v}(-\theta_j-1),\no\\[4pt]
&&\hat{T}_1^{  v}(-\theta_j)\hat{\bar T}_{\langle23\rangle}^{
v}(-\theta_j-1)=P_{123}\hat{T}_1^{
v}(-\theta_j)\hat{\bar T}_{\langle23\rangle}^{  v}(-\theta_j-1),\no\\[4pt]
&&\hat{T}_1^{  v}(-\theta_j)\hat{\bar T}_{\langle234\rangle}^{
v}(-\theta_j-1)=P_{1234}\hat{T}_1^{
v}(-\theta_j)\hat{\bar T}_{\langle234\rangle}^{  v}(-\theta_j-1),\no\\[4pt]
&&\hat{T}_2^{  v}(-\theta_j)\hat{T}_{\bar 1}^{
\bar{v}}(-\theta_j-1)=P^{{  v\bar{v}}(5)
}_{2\bar{1}}\hat{T}_2^{  v}(-\theta_j)\hat{T}_{\bar 1}^{
\bar{v}}(-\theta_j-1),\no\\[4pt]
&&\hat{T}_2^{  v}(-\theta_j)\hat{T}_{\tilde 1}^{
\tilde{v}}(-\theta_j-\frac{1}{{2}})=P^{{  v\tilde{v}}(11)
}_{2\tilde{1}}\hat{T}_2^{  v}(-\theta_j)\hat{T}_{\tilde 1}^{
\tilde{v}}(-\theta_j-\frac{1}{{2}}),\no\\[4pt]
&&\hat{T}_2^{  v}(-\theta_j)\hat{T}_{ 1}^{
s}(-\theta_j-\frac{5}{{4}})=P^{{  vs}(4) }_{21}\hat{T}_2^{
v}(-\theta_j)\hat{T}_{ 1}^{
s}(-\theta_j-\frac{5}{{4}}).\label{fuii-7} \eea

%%%%%%%%%%%%%%%%%%%%%%%%%%%%%%%%%%%%%%%%%%%%%%%%%%%%%%%%%%%%%%%
%                                                             %
%  References                                                 %
%                                                             %
%%%%%%%%%%%%%%%%%%%%%%%%%%%%%%%%%%%%%%%%%%%%%%%%%%%%%%%%%%%%%%%

\end{document}